\documentclass[aps,superscriptaddress,twocolumn,showpacs]{revtex4}

\pdfoutput=1

\usepackage{amsmath,amssymb,eucal,graphicx}

\usepackage[dvipsnames]{xcolor}

\begin{document}

\title{Power spectral density of a single Brownian trajectory: What one can
and cannot learn from it}

\author{Diego Krapf}
\affiliation{Department of Electrical and Computer Engineering, Colorado State
University, Fort Collins, Colorado 80523, USA}
\affiliation{School of Biomedical Engineering, Colorado State University, Fort
Collins, Colorado 80523, USA}
\author{Enzo Marinari}
\affiliation{Dipartimento di Fisica, Sapienza Universit{\`a} di Roma, P.le A.
Moro 2, I-00185 Roma, Italy}
\affiliation{INFN, Sezione di Roma 1 and Nanotech-CNR, UOS di Roma, P.le A.
Moro 2, I-00185 Roma, Italy}
\author{Ralf Metzler}
\affiliation{Institute for Physics \& Astronomy, University of Potsdam,
Karl-Liebknecht-Str 24/25, D-14476 Potsdam-Golm, Germany}
\author{Gleb Oshanin}
\affiliation{Laboratoire de Physique Th\'{e}orique de la Mati\`{e}re Condens\'{e}e 
(CNRS UMR 7600), Sorbonne Universit\'{e}, 4 Place Jussieu, 75252 Paris Cedex
05, France}
\author{Xinran Xu}
\affiliation{Department of Electrical and Computer Engineering, Colorado State
University, Fort Collins, Colorado 80523, USA}
\author{Alessio Squarcini}
\affiliation{Max-Planck-Institut f\"ur Intelligente Systeme, Heisenbergstr. 3,
D-70569, Stuttgart, Germany}
\affiliation{IV. Institut f\"ur Theoretische Physik, Universit\"at Stuttgart,
Pfaffenwaldring 57, D-70569 Stuttgart, Germany}

\begin{abstract}
The power spectral density (PSD) of any time-dependent stochastic processes
$X_t$ is a meaningful feature of its spectral content. In its text-book
definition, the PSD is the Fourier transform of the covariance function
of $X_t$ over an infinitely large observation time $T$, that is, it is
defined as an ensemble-averaged property taken in the limit $T \to \infty$.
A legitimate question is what information on the PSD can be reliably obtained
from single-trajectory experiments, if one goes beyond the standard definition
and analyzes the PSD of a \textit{single} trajectory recorded for a \textit{finite} observation
time $T$.  In quest for this answer, for a $d$-dimensional Brownian motion
we calculate the probability density function of a single-trajectory PSD for
arbitrary frequency $f$, finite observation time $T$ and arbitrary number
$k$ of projections of the trajectory on different axes. We show analytically
that the scaling exponent for the frequency-dependence of the PSD specific
to an ensemble of Brownian motion trajectories can be already obtained from
a single trajectory, while the numerical amplitude in the relation between
the ensemble-averaged and single-trajectory PSDs is a fluctuating property
which varies from realization to realization.  The distribution of this
amplitude is calculated exactly and is discussed in detail. Our results are
confirmed by numerical simulations and single particle tracking experiments,
with remarkably good agreement.  In addition we consider a truncated
Wiener representation of Brownian motion, and the case of a discrete-time
lattice random walk.  We highlight some differences in the behavior of a
single-trajectory PSD for Brownian motion and for the two latter situations.
The framework developed herein will allow for meaningful physical analysis
of experimental stochastic trajectories.
\end{abstract}


\maketitle

\section{Introduction}

The power spectral density (PSD) of a stochastic process $X_t$, which is formally
defined as (see, e.g., Ref.\cite{def})
\begin{align}
\label{spec0}
\mu_S(f,\infty)=\lim_{T\to\infty}\frac{1}{T}\mathbb{E}\left\{\left|\int^T_0
e^{i f t}X_t dt\right|^2\right\},
\end{align}
provides important insights into the spectral content of $X_t$. Here and in
what follows, the symbol  $\mathbb{E}\left\{ \ldots\right\}$ denotes averaging
with respect to all possible realizations of the process, i. e., the expectation.  
Often the PSD
as defined in Eq. \eqref{spec0} has the form $\mu_S(f,\infty) = A/f^{\beta}$,
where $A$ is an amplitude and $\beta$, (typically, one has $0 < \beta \leq 2$ \cite{mike}), is the exponent characteristic of the
statistical properties of $X_t$.  In experiments and numerical modeling,
the PSD has been determined using a periodogram estimate for a wide variety
of systems in physics, biophysics, geology etc.  A by far non-exhaustive list  of systems exhibiting $1/f^{\beta}$ spectra
includes electrical signals in vacuum tubes, semiconductor devices, metal films
and other condensed matter systems \cite{1,2,3},  quantum dots \cite{4,5,14},
nano-electrodes \cite{6} and two-dimensional graphene layers with a widely
 tunable concentration of carriers \cite{69}. The PSD has been analysed, as well, for the trajectories of tracers in artificial crowded fluids \cite{weiss},
 for active 
 micro-rheology of colloidal suspensions
\cite{66}, Kardar-Parisi-Zhang interface fluctuations \cite{77}, for sequences of earthquakes \cite{7}, weather data \cite{8,talk1,talk2},
biological evolution \cite{9}, human cognition \cite{10}, network traffic \cite{11} and even for the loudness of music recordings (see, e.g., Refs. \cite{777,888}).

On the theoretical side, the PSDs showing the $1/f^{\beta}$ dependences have been calculated analytically for
diverse situations, including, e.g., the dynamics in chaotic Hamiltonian systems \cite{999}, 
periodically-driven bistable systems \cite{han}, fluctuations of the phase-separating interface  \cite{ale},
several 
diffusive and non-diffusive transport processes
(see, e.g., Refs. \cite{14,15,150,16,17}),  the running maximum of Brownian
motion \cite{29}, diffusion in presence of strong quenched disorder
\cite{30,31,32,33} and 
the electric-field-driven ion transport through nanometer-scale membrane
pores \cite{333}. The PSD of Brownian motion in an optical trap has been
scrutinized in Refs. \cite{flyv,sch}, which permitted the calibration of
optical tweezers, making them a powerful tool for force spectroscopy, local
viscometry, and other applications (see, e.g. Ref. \cite{chris,lenerev}).

The systematic measurements of single particle trajectories started with Perrin more than a century ago \cite{perrin}.
Just a few years later Nordlund \cite{nordlund} developed impressive experimental
techniques,  followed by a line of further refinements up to Kappler's
measurements \cite{kappler}, to record sufficiently long trajectories to enable
quantitative analysis on the basis of individual trajectories---without
the need of prior ensemble averaging.
Nowadays, with the advent of modern
microscopy and supercomputing, scientists routinely measure long trajectories 
$X_t$ of submicron tracer particles or even single molecules 
\cite{weiss,phystoday,tak,18,181,19,lenerev}.

In parallel to this experimental progress, 
there was a general shift  of  interest towards
understanding statistical properties of individual realizations of stochastic
processes.  
In particular, a conceptually important question often raised
within this context is if one can reliably extract information about the
ensemble-averaged properties of random processes from single-trajectory data
\cite{phystoday}. 
Considerable theoretical progress has been achieved, for instance,
in finding the way to get the ensemble-averaged diffusion coefficient
from a single Brownian trajectory, which task amounts to seeking properly
defined functionals of the trajectory which possess an ergodic property (see,
e. g., Refs. \cite{18,181,19,phystoday,20,201,2011,21,210,22,23,24,norre} for a general overview).

One of such functionals
used in single-trajectory analysis 
is the time averaged mean squared displacement (MSD)
(see, e.g., Refs.\cite{18,181,19,phystoday,lenerev,tak})
\begin{align}
\label{tamsd}
\overline{\delta^2(\tau)}=\frac{1}{T-\tau}\int_0^{T-\tau}\Big(X_{t+\tau}
- X_t\Big)^2dt,
\end{align}
where the bar denotes time averaging and 
the lag time $\tau$ acts as the size of the analysis window sliding
over the time series $X_t$. For finite observation time $T$, the time averaged
MSD varies randomly  for different trajectories
even under identical physical conditions. For both normal
and anomalous diffusion this variation is mainly seen as an
amplitude scatter of $\overline{\delta^2(\tau)}$ at given lag times, which  
remains remarkably constant (apart of relatively small local fluctuations)
as function of $\tau$ \cite{pccp,thiel,jeon,ghosh}.

For the time averaged MSD the fluctuations are often quantified in terms of the
so-called ergodicity breaking parameter $\mathrm{EB}=\mathbb{E}\{\xi^2\}-1$, where
the dimensionless variable $\xi=\overline{\delta^2(\tau)}/\mathbb{E}\{\overline{
\delta^2(\tau)}\}$ for a given $\tau$ measures the deviations from the
trajectory-average $\mathbb{E}\{\overline{\delta^2(\tau)}\}=(1/N)\sum_{
i=1}^N\overline{\delta^2_i(\tau)}$ \cite{he,pccp}, with $N$ being the number of
observed trajectories. Clearly, $\mathrm{EB} \equiv \gamma^2$ 
with $\gamma$ being
the coefficient 
of variation of the distribution $P(\delta^2(\tau))$.

For Brownian motion, the fluctuations decrease with growing
observation time such that $\mathrm{EB}\sim 4\tau/(3 T)$ \cite{181,2011,pccp}, which
permits to deduce the diffusion coefficient specific to an ensemble of trajectories
from just a single, sufficiently long, trajectory.
A similar decay of EB to zero for large $T$ is observed for
fractional Brownian motion \cite{deng} and scaled Brownian motion \cite{sbm}.
However, for anomalous diffusion with scale-free distributions of waiting times
\cite{he} and processes with systematic, spatially varying diffusion coefficient
\cite{hdp} the inequality $\mathrm{EB} \neq 0$ persists even in the limit $T\to
\infty$, which reveals  ageing in the sense that the properties of the system,
such as the effective diffusivity, are perpetually changing during the measurement
and depend on $T$ \cite{phystoday,pccp,johannes}. In these systems, 
the fluctuations as measured by $\mathrm{EB}$ do not decay to zero with increasing $T$,
and thus time averages of physical observables remain random quantities,
albeit with well defined distributions (see, e.g. Refs. \cite{2011,he,pccp}).  
Many other facets of the time averaged MSD in Eq. \eqref{tamsd} and of 
the parameter $\mathrm{EB}$
  were 
exhaustively 
well studied to date for a variety of 
diffusive processes,  providing a solid mathematical framework for the analysis of individual trajectories 
 \cite{pccp,lenerev,igorrev,hoefling,deng,maria,bau}.

Single-trajectory PSDs have been 
studied 
in several cases. In particular, the PSD of the loudness of musical recordings was discussed in Refs. \cite{777,888} and the spectra of temperature data were presented in Refs. \cite{talk1,talk2}. 
We note that, of course, for these two examples 
averaging over "an ensemble" does not make any sense since the latter simply 
does not exist, likewise, e.g., the case of the financial markets data for which the single-trajectory MSD has been recently analyzed \cite{money}.  Further on,
 power spectra of individual time-series  
 were examined for a two-state stochastic model describing blinking quantum dots 
\cite{15} and also for single-particle tracking experiments with tracers in artificially crowded fluids \cite{weiss}.
It was shown that, surprisingly enough, the estimation of the exponent characterising the power spectrum
of an ensemble from the single-trajectory PSD is rather robust.  Next, in 
Refs. \cite{ameb,ameb2}
the power spectra of the velocities of  independent
motile amoeba 
(see also Refs. \cite{tak,bod}) were analysed, revealing a robust 
high-frequency 
asymptotic $1/f^2$  form, 
persisting for all recorded individual trajectories. 

Clearly, these 
numerical 
observations raise several challenging questions. 
Indeed, could it be possible that the exponent characterising the 
frequency-dependence of the standard PSD, defined as an ensemble-average property, 
can be observed already on 
a single-trajectory level? 
If true, does it hold for any stochastic transport process or just for some particular examples?
Moreover, in which range of frequencies can such a behavior be observed? 
What are the distributions of the amplitudes entering the relation between a single-trajectory PSD and its ensemble-average 
counterpart and how broad are they?
Evidently, 
numerical analyses may give a certain degree of understanding of some particular features, but 
a deep insight can be obtained  only in conjunction with a full mathematical solution.
Unlike the single-trajectory MSD, for which a deep knowledge has been already acquired via 
an exact analytical analysis, a similar analysis 
of the statistical properties of a single-trajectory PSD is lacking at present, although 
 its ensemble-averaged counterpart is widely
used as an important quantifier of different properties of random trajectories
in diverse areas of engineering, physics and chemistry. 

In this paper, going beyond the text-book definition \eqref{spec0}, we
concentrate on the question what information can be reliably obtained if
one defines the PSD of a single, finite-time realization of $X_t$. 
We here focus on the paradigmatic process of Brownian motion (BM).
This choice is two-fold: first, Brownian motion is ubiquitous in nature which renders this analysis particularly important.
Second, it permits us 
to obtain an exact mathematical solution of the problem: we calculate
exactly the moment-generating function and the full probability density
function of the single-trajectory PSD and its moments of arbitrary order in
the most general case of arbitrary frequency, arbitrary (finite) observation
time and arbitrary number of the projections of a $d$-dimensional BM onto
the coordinate axes. This furnishes yet another example of a time-averaged
 functional of BM, whose moment-generating function and full
probability distribution can be calculated exactly (see, e. g.,
Ref. \cite{satya}). 

Capitalizing on these results, we observe that for a sufficiently large
$T$ (and frequency $f$ bounded away from zero) for any realization of the
process a single-trajectory PSD is \emph{proportional\/} to its first moment
\eqref{spec0}, and the latter embodies the full dependence on the frequency
and on the diffusion coefficient specific to an ensemble of the trajectories.
This means that \emph{the frequency-dependence can be deduced
from a single trajectory}. In other words, there is no need to perform
averaging over an ensemble of trajectories---one long trajectory suffices.
However, the proportionality factor, connecting a single-trajectory PSD and its
ensemble-averaged counterpart---a numerical amplitude---is random and varies
from realization to realization. Due to this fact, one cannot infer the value
of the ensemble-averaged diffusion coefficient from the amplitude of a single-trajectory PSD.
The distribution function of this amplitude is calculated exactly here and
its effective width is quantified using standard criteria.

As a proof of concept, we revisit our predictions
for a continuous-time BM---an idealized process with infinitesimal
increments---resorting to a numerical analysis based on Monte Carlo simulations
of discrete-time random walks, and also using  experimental single-trajectory
data for the diffusive motion of micron-sized polystyrene beads in a flow cell.
We demonstrate that our theoretical prediction for the relation connecting
a single-trajectory PSD and its ensemble-averaged counterpart is
corroborated by numerical and experimental results. Additionally we show
that the predicted distribution of the amplitude is consistent with
numerical results, which means that the framework developed here is 
justified and allows for a meaningful analysis of experimental trajectories.

Pursuing this issue further, we address several general questions emerging
in connection with a comparison of our analytical predictions against
numerical simulations and experimental data. To this end, we first consider
Wiener's  representation of BM in form of an infinite Fourier series with
random coefficients, whose truncated version is often used  in numerical
simulations. We show that the distribution of the single-trajectory PSD
obtained from Wiener's representation in which just $N$ terms are kept,
instead of an infinite number, has exactly the same form as the one obtained
for the continuous-time BM when  $f T$ is within the interval $(0, \pi N)$.
Outside of this interval, the probability density function of the truncated
PSD converges to a different form.

We examine the case when a trajectory of a continuous-time BM is recorded
at some discrete time moments, so that it is represented
by a set of $M$ points. The single-trajectory PSD, i.e., the periodogram,
becomes a periodic function of $f T$ with the prime period equal to $2 \pi M$.
We analyze several aspects of this discrete-time problem: We study
how large $M$ should be taken at a fixed observation time $T$ so that we
may recover the results obtained for the continuous-time BM. We analyze the
limiting forms of the distribution of a single-trajectory periodogram and
show, in particular, that for $f T$ kept fixed and $M \to \infty$, the distribution
converges to the form obtained for a continuous-time BM. On the contrary,
when $f T$ is left arbitrary so that it may assume any value within the
prime period, that is, $f T \in (0, 2 \pi M)$, the distribution
of  a single-trajectory periodogram converges to a different limiting form as $M
\to \infty$. Our analysis demonstrates that when $fT$ belongs to a certain
interval within the prime period, a single-trajectory periodogram equals,
up to a random numerical amplitude, the ensemble-averaged periodogram, and
the latter embodies the full dependence on $f$ and $T$. Therefore, similarly
to the continuous-time case, the correct spectrum can be obtained already
from a single trajectory.
These new results are the foundation for the use of single
trajectory PSD in the quantitative analysis of single or few recorded particle
trajectories, complementing the widely used concept of the single trajectory MSD.

The paper is outlined as follows: In Sec. \ref{BM1} we introduce our basic
notations. In Sec. \ref{BM} we first derive explicit expressions for the
variance of a single-trajectory PSD and the corresponding coefficient of
variation of the probability density function, and present exact results for
the moment-generating function of a single-trajectory PSD, its full probability
density function and moments of arbitrary order (Sec. \ref{momgen}).
Section \ref{ST} is devoted to the relation connecting a single-trajectory
PSD and its ensemble-averaged counterpart, while Sec.  \ref{FL} discusses
fluctuations of the amplitude in this relation. Next, in Sec. \ref{TR}
we analyze the probability density function of a single-trajectory PSD
obtained by truncating Wiener's representation of a continuous-time BM.
Section \ref{discrete} presents an analogous analysis for the case when
a continuous-time trajectory is recorded at discrete time moments. In
Sec. \ref{conc} we conclude with a brief recapitulation of our results and
outline some perspectives for further research.  Additional details are
relegated to Appendix \ref{A}, in which we present exact results for the
distributions in the special case $f=0$, and to Appendix \ref{B}, where we discuss several
cases in which the full distribution of a single-trajectory periodogram can
be evaluated exactly in the discrete-time settings.

\section{Brownian motion: Definitions and notations}
\label{BM1}

Let ${\bf X}_t$, $t \in (0,T)$ denote a Brownian trajectory in a $d$-dimensional
continuum and $X_t^{(j)}$ with $j=1,2, \ldots, d$ stand for the projection of
${\bf X}_t$ on the axis $x_j$. The projections $X_t^{(j)}$ are statistically
independent of each other and (likewise the BM ${\bf X}_t$ itself) each projection 
$X_t^{(j)}$ is a Gaussian process with zero mean and variance
\begin{align} 
\mathbb{E}\left\{\left(X_t^{(j)}\right)^2\right\}=2Dt,
\end{align}
where $D$ is the diffusion coefficient of ${\bf X}_t$. In a random walk sense, 
$D=\langle\delta x^2\rangle/(2d\langle\delta\tau\rangle)$, where $\langle\delta
x^2\rangle$ is the variance of the step length and $\langle\delta\tau\rangle$ is
the mean time interval between consecutive steps. Hence, $D$ contains the dimension
$d$ of the unprojected motion. 

In text-book notations, the PSD of each of the projections is defined by
Eq.~\eqref{spec0}, that is,  
\begin{align}
\label{spec1}
&\mu_S^{(j)}(f,\infty) =  \lim_{T \to \infty} \mu^{(j)}_S(f,T) \,,
\end{align} 
where (taking into account that $X_t^{(j)}$  is real-valued)  
the $T$-dependent function  $\mu_S^{(j)}(f,T)$ 
is given explicitly by
\begin{eqnarray}
\label{text-book}
\mu^{(j)}_S(f,T)&=&\frac{1}{T}\int^T_0\int^T_0dt_1dt_2\cos\left(f\left(t_1-
t_2\right)\right)\nonumber\\
&&\times\mathbb{E}\left\{X^{(j)}_{t_1} X^{(j)}_{t_2}\right\},
\end{eqnarray}
$\mathbb{E}\left\{X^{(j)}_{t_1} X^{(j)}_{t_2}\right\}=2D \, {\rm min}(t_1,t_2)$ 
being the covariance of $X_t^{(j)}$. For BM, one then finds the standard result
\cite{def}
\begin{align}
\label{power_av}
\mu_S^{(j)}(f,T)=\frac{4D}{f^2}\left[1-\frac{\sin\left(f T\right)}{f T}\right],
\end{align}
such that
\begin{align}
\label{xxx}
\mu_S^{(j)}(f,\infty) = \frac{4 D}{f^2} \,.
\end{align}
Hence, for BM the standard power spectral density $\mu_S^{(j)}(f,\infty) $, defined as an average over an ensemble of trajectories,  is described by
 a power-law with characteristic exponent $\beta = 2$ and 
an amplitude which is  linearly proportional to the diffusion coefficient $D$.

Going beyond the text-book definition in Eq. \eqref{text-book}, we now define the PSD of a single component $X^{(j)}_t$:
\begin{align}
\label{single}
S^{(j)}_{T}(f)   =\frac{1}{T} &\int^T_0 \int^T_0 dt_1 dt_2 \cos\left(f \left(t_1 - t_2\right)\right) X^{(j)}_{t_1} X^{(j)}_{t_2} \,,
\end{align}
and another property of interest here, the \emph{partial\/} PSD of the trajectory
${\bf X}_t $
\begin{eqnarray}
\label{partial}
\tilde{S}^{(k)}_{T}(f)&=&\frac{1}{T}\int^T_0\int^T_0dt_1dt_2\cos\left(f\left(t_1
-t_2\right)\right)\nonumber\\
&&\hspace*{-1.2cm}\times\Bigg[X^{(1)}_{t_1} X^{(1)}_{t_2}+X^{(2)}_{t_1}X^{(2)}_{
t_2}+\ldots+X^{(k)}_{t_1} X^{(k)}_{t_2} \Bigg],
\end{eqnarray}
in which we take into account the contributions of $k$, ($k=1,2, \ldots,d$),
components of a $d$-dimensional BM. Clearly, for $k =1$ the definitions
\eqref{single} and \eqref{partial} coincide. The PSDs $S^{(j)}_{T}(f)$ and
$\tilde{S}^{(k)}_{T}(f)$ are $f$- and $T$-parameterized random variables:
the first moment $\mu_S(f,T)$ of a single-trajectory PSD  \eqref{single}
is given by the standard result  \eqref{power_av}, while the first moment
of the partial PSD  \eqref{partial}, due to statistical independence of the
components,  is given by expression \eqref{power_av} multiplied by  $k$.
Our goal is to evaluate exactly  the full probability density function for
$\tilde{S}^{(k)}_{T}(f)$.

\section{Brownian motion: Results}
\label{BM}

\begin{figure}
\includegraphics[width=8.5cm]{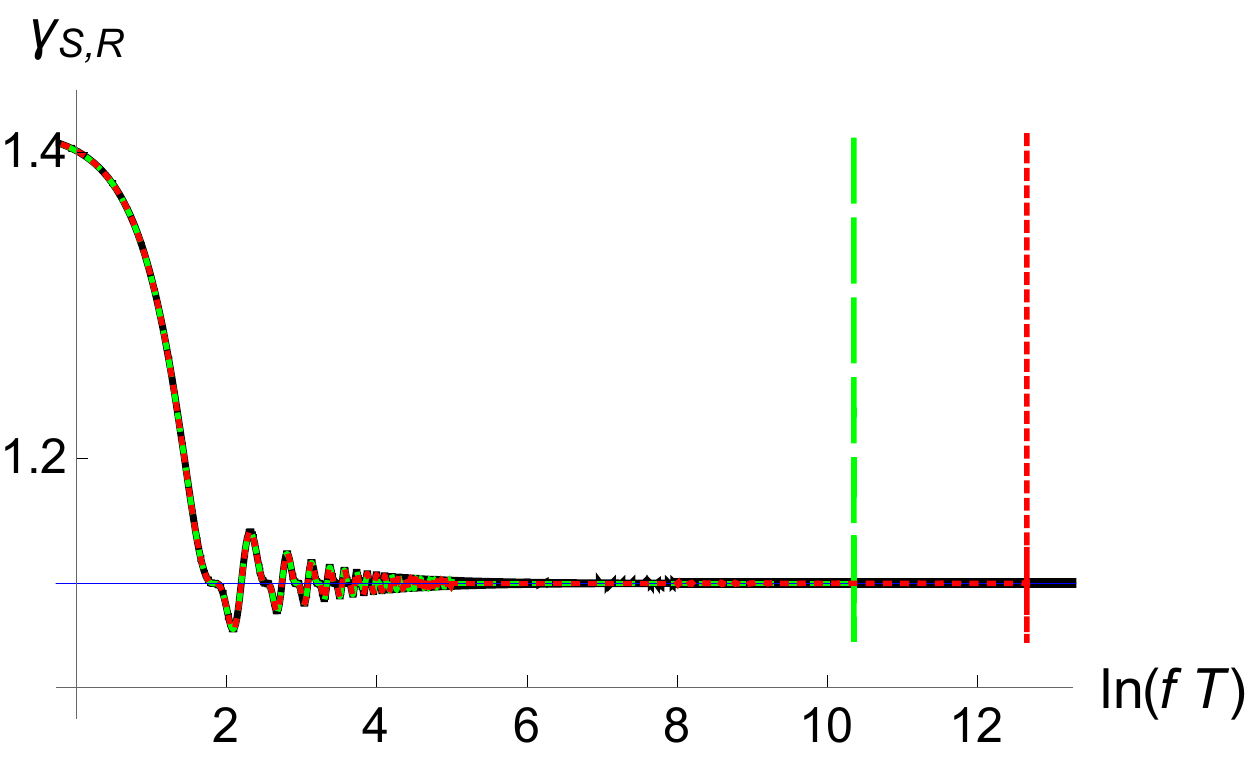}
\caption{Coefficient of variation $\gamma$ as function of
$\ln(f T)$. The solid curve is the analytical expression for continuous-time
BM, $\gamma_S = \sigma_S(f,T)/\mu_S(f,T)$ (see Eqs. \eqref{power_av}
and \eqref{var}).  The thin horizontal line (blue) is the asymptotic value
$\sqrt{5}/2$ attained by $\gamma_S$ in the limit $f T \to \infty$. The
dashed (green) and dotted (red) curves show the coefficient of variation
$\gamma_R = \sigma_R(f,T)/\mu_R(f,T)$ in the discrete-time case  (see
Sec. \ref{discrete}, Eqs. \eqref{dis_mean} and \eqref{dis_var}) with
the number $M$ of the recorded positions of a trajectory equal to $10^4$
and $10^5$, respectively.  The vertical peaks of $\gamma_R$ (of height $=
\sqrt{2}$) appear at $f T = \pi M$, (that is, in the middle of the prime period),
corresponding to $f T = \pi \times 10^4$ (green) and $f T = \pi \times 10^5$
(red). Note that the oscillations of $\gamma$ apparent for small and moderate
values of $f T$ also exist for the discrete-time case close to the "peaks"
but are not resolved in the figure due to the logarithmic scale.  For the
discrete-time case on a linear scale $\gamma_R$ has a periodic term in $f T$
(see Sec. \ref{discrete} and Fig. \ref{sketch} for more details). \label{FIG1}}
\end{figure}

To get an idea of how representative of the actual behavior of a
single-trajectory PSD the result \eqref{power_av} is, we first look at the
variance $\sigma_S^2(f,T)$ of a single-component single-trajectory PSD,
\begin{eqnarray}
\label{var}
\sigma_S^2(f,T)&=&\mathbb{E}\left\{\left(S^{(j)}_T(f)\right)^2\right\}-\mu_S^2
(f,T)\nonumber\\
&=&\frac{20 D^2}{f^4}\Bigg[1-\Big(6-\cos\left(f T\right)\Big)\frac{2\sin\left(
fT\right)}{5fT}\nonumber\\
&&+\frac{\Big(17-\cos\left(2fT\right)-16\cos\left(f T\right)\Big)}{10 f^2 T^2}
\Bigg].
\end{eqnarray}
Expression \eqref{var} permits us to determine the corresponding coefficient
of variation 
\begin{equation}
\label{coeff_var}
\gamma_S = \sigma_S(f,T)/\mu_S(f,T)
\end{equation}
of the yet unknown distribution of a single-component single-trajectory PSD.

In Fig.~\ref{FIG1} we depict $\gamma_S$, which is a function of the product
$f T$ exclusively.  We observe that $\gamma_S$ approaches $\sqrt{2}$ for
fixed $T$ and $f \to 0$, and tends to the asymptotic value $\sqrt{5}/2$
(thin horizontal line in Fig. \ref{FIG1}) when $T \to \infty$ at any fixed $f > 0$.
Overall, for any value of $f T$ the coefficient of variation appears to
be greater than unity, meaning that the standard deviation $\sigma_S(f,T)$
is greater than the average, $\mu_S(f,T)$, which signals that the parental
distribution  is effectively "broad", despite the fact that it evidently
has well-defined moments of arbitrary order.  As a consequence, the average
described by Eq. \eqref{power_av} may indeed not be representative of the
actual behavior of a single-trajectory PSD. This fully validates our quest for
the distribution $P\left(\tilde{S}^{(k=1)}_{T}(f)\right)$.

\subsection{Brownian motion: moment-generating-function, distribution
$P\left(\tilde{S}^{(k)}_{T}(f)\right)$ and its moments.}
\label{momgen}

Our first goal is 
to calculate the moment-generating function of $\tilde{S}^{(k)}_{T}(f)$ in Eq. \eqref{partial}, defined formally as the  following Laplace transform
\begin{align}
\label{Phi}
\Phi_{\lambda} = \Phi_{\lambda}\left(\tilde{S}^{(k)}_{T}(f)\right) \equiv \mathbb{E}\left\{ \exp\left(- \lambda \, \tilde{S}^{(k)}_{T}(f) \right) \right\} \,.
\end{align}
To calculate $\Phi_{\lambda}$ exactly, it appears  
convenient to use  Wiener's representation of  
a given Brownian path
$X^{(j)}_t$ in the form of a Fourier series with random coefficients 
\begin{align}
\label{wiener}
X^{(j)}_t = \frac{\sqrt{2 T}}{\pi} \sum_{n=1}^{\infty} \frac{\zeta^{(j)}_n}{(n - 1/2)} \sin\left(\frac{(n - 1/2) \pi t}{T}\right) \,,
\end{align}
where $\zeta^{(j)}_n$ are independent, identically-distributed random variables with the distribution
\begin{align}
\label{zeta}
P\left(\zeta^{(j)}_n\right) = \frac{1}{\sqrt{4 \pi D}} \exp\left(- \left(\zeta^{(j)}_n\right)^2/4 D\right) \,.
\end{align}

The corresponding single-trajectory partial PSD $\tilde{S}^{(k)}_T(f)$ in
Eq.~\eqref{partial} can be formally rewritten as 
\begin{align}
\label{def2}
\tilde{S}^{(k)}_T(f) = \frac{2}{\pi^2} \sum_{j=1}^k \left[ \left(\sum_{n=1}^{\infty} g_n \zeta^{(j)}_n\right)^2 + \left(\sum_{n=1}^{\infty} h_n \zeta^{(j)}_n\right)^2 \right]\,, 
\end{align}
where the functions $g_n=g_n(f,T)$ and $h_n=h_n(f,T)$ are given explicitly by
\setcounter{equation}{15}
\begin{subequations}
\begin{align}
g_n=\frac{T}{(n-1/2)}\frac{(-1)^nfT\sin\left(fT\right)+\pi(n-1/2)}{\left(\pi^2
(n-1/2)^2-(fT)^2\right)},
\end{align}
and
\begin{align}
h_n=-\frac{T}{(n-1/2)}\frac{(-1)^nfT\cos\left(fT\right)}{\left(\pi^2(n-1/2)^2-
(fT)^2\right)}.
\end{align}
\end{subequations}
Inserting expression \eqref{def2} into \eqref{Phi} and using 
the 
identity
\begin{align}
\label{identity}
\exp\left(- \frac{b^2}{4 a}\right) = \sqrt{\frac{a}{\pi}} \int^{\infty}_{-\infty} dx \exp\left(- a x^2 + i b x\right) \,, 
\end{align}
we rewrite Eq.~\eqref{Phi} in the factorized form
\begin{widetext}
\begin{align}
\Phi_{\lambda} &= \Bigg[\frac{\pi}{8 \lambda} \int^{\infty}_{-\infty} \int^{\infty}_{-\infty} dx \, dy \exp\left(- \frac{\pi^2}{8 \lambda} \left(x^2 + y^2\right)\right) \mathbb{E}_{\zeta}\left\{\exp\left( i \sum_{n=1}^{\infty} \left(x g_n + y h_n\right) \zeta^{(j)}_n\right) \right\}\Bigg]^k \,,
\end{align}
\end{widetext}
where the subscript $\zeta$ in the averaging operator $\mathbb{E}_{\zeta}\{\ldots \}$ signifies that averaging over all paths of the component $X_t^{(j)}$ is replaced by an equivalent operation, the averaging over all possible values of $\zeta_n^{(j)}$. Performing this averaging, as well as the integrations over $dx$ and $dy$, we get
\begin{align}
\label{sums}
\Phi_{\lambda} =&\Bigg[1 + \frac{8 \lambda D}{\pi^2} \sum_{n=1}^{\infty}\left(g_n^2 + h_n^2\right) + \left(\frac{8 \lambda D}{\pi^2}\right)^2 \nonumber\\
& \times \Bigg(\sum_{n,l=1}^{\infty} g_n^2 h_l^2 - \left(\sum_{n=1}^{\infty} g_n h_n\right)^2\Bigg)\Bigg]^{-k/2} \,.
\end{align}
We focus 
next on the infinite sums entering Eq. \eqref{sums}. They can be calculated exactly, and 
 expressed via the first and the second moments of a single-trajectory PSD, 
\setcounter{equation}{19}
\begin{subequations}
\begin{align}
\sum_{n=1}^{\infty}\left(g_n^2 + h_n^2\right) \equiv \frac{\pi^2}{4 D} \mu_S(f,T)
\end{align}
and 
\begin{align}
&\sum_{n,l=1}^{\infty} g_n^2 h_l^2 - \left(\sum_{n=1}^{\infty} g_n h_n\right)^2  \equiv 
 \left(\frac{\pi^2}{8 D}\right)^2 \mu_S^2(f,T)  \left(2 - \gamma_S^2
\right) \,,
\end{align}
\end{subequations}
where the average and the variance (as well as the corresponding coefficient $\gamma_S$ of variation)  of the single-component single-trajectory PSD 
are defined in Eqs.~\eqref{power_av}, \eqref{var}, and \eqref{coeff_var}, 
respectively. 

Consequently we realize that the moment-generating function in Eq.~\eqref{sums} 
can be cast into a more compact and physically meaningful form, which involves
only the first moment and the variance (through the coefficient $\gamma_S$ of
variation) of the single-trajectory PSD,
\begin{align}
\label{MG}
\Phi_{\lambda} &= \Bigg[1 + 2  \, \mu_S(f,T) \, \lambda \,+ 
 \left(2  - \gamma_S^2\right) \mu_S^2(f,T) \, \lambda^2\Bigg]^{-k/2} \,.
\end{align} 
This expression holds for any value of $k$, $f$ and $T$. 

Performing next the inverse Laplace transform of the function defined by Eq. \eqref{MG}, 
we arrive at the 
following expression for the desired probability density function of the single-trajectory \textit{partial} PSD defined in Eq. \eqref{partial}, which also (as the result in Eq. \eqref{MG}) holds for arbitrary $f$, arbitrary  $T$ and arbitrary number $k$ of the projections of the trajectory ${\bf X}_t$ onto the coordinate axes,
\begin{widetext}
\begin{align}
\label{dist}
P(\tilde{S}_T^{(k)}(f)=S) &= \frac{\sqrt{\pi}}{2^{\frac{k-1}{2}} 
\Gamma\left(k/2\right)} \frac{S^{\frac{k-1}{2}}}{\sqrt{2 - \gamma_S^2} \left(\gamma_S^2 -1 \right)^{\frac{k-1}{4}} \mu_S^{\frac{k+1}{2}}(f,T)}  \exp\left(- \frac{1}{2 - \gamma_S^2} \frac{S}{\mu_S(f,T)}\right) {\rm I}_{\frac{k-1}{2}}\left(\frac{\sqrt{\gamma_S^2 - 1}}{2 - \gamma_S^2} \frac{S}{\mu_S(f,T)}\right).
\end{align} 
\end{widetext}
Here, $I_{\alpha}(\ldots)$ is the modified Bessel function of the $1$st kind and $\Gamma(\ldots)$ is the Gamma-function. 

Before we proceed further, two remarks are in order. First, we note that the
distribution \eqref{dist} is the Bessel function distribution that has been
used in mathematical statistics years ago  as an example of a distribution with
ÒheavierÓ than Gaussian tails (see, e.g., Refs. \cite{math2,math1}). Second,
as already mentioned, for $f = 0$ the coefficient of variation is exactly
equal to $\sqrt{2}$ and hence, the coefficient in front of the term quadratic
in $\lambda$ in expression \eqref{MG} vanishes. In this particular case the
distribution \eqref{dist} simplifies to the $\chi^2$-distribution
with $k$ degrees of freedom, which is presented in Appendix \ref{A}.

Expression \eqref{dist} permits us to straightforwardly calculate the moments
of the partial single-trajectory PSD of arbitrary, not necessarily integer order
$Q>-(k+1)/2$,
\begin{eqnarray}
\label{moments}
\frac{\mathbb{E}\left\{\left(\tilde{S}_T^{(k)}(f)\right)^Q\right\}}{\mu^Q_S(f,T)}
&=&\frac{\Gamma\left(Q+k\right)\left(2-\gamma_S^2\right)^{Q+k/2}}{\Gamma\left(k
\right)}\nonumber\\
&&\hspace*{-2.8cm}\times \,_2F_1\left(\frac{Q+k}{2},\frac{Q+k+1}{2};\frac{k+1}{2};
\gamma^2_S-1\right),
\end{eqnarray}
where $_2F_1\left(\dots\right)$ is the Gauss hypergeometric function. Note that the moments of order $Q > 2$ for an arbitrary $k$
are all expressed through the first and the second (via $\gamma_S$) moments of the single-component single-trajectory PSD only, 
since the parental Gaussian process $X_t^{(j)}$ is entirely defined by its first two moments. 

The distribution in Eq. \eqref{dist} and the formula for the moments of order $Q$, Eq. \eqref{moments}, ensure that the single-trajectory PSD $\tilde{S}_T^{(k)}(f)$ has the form
\begin{align}
\label{conc0}
\tilde{S}_T^{(k)}(f) = A^{(k)}\left(\gamma_S\right) \, \mu_S(f,T) \,,
\end{align}
where $\mu_S(f,T)$ is the first moment of this random variable, i.e., a deterministic function of $f$ and $T$ 
which sets the scale of variation of  $\tilde{S}_T^{(k)}(f)$, and $A^{(k)}\left(\gamma_S\right)$ is a \textit{dimensionless} random amplitude, whose moments of order $Q$ are defined by the expression in the right-hand-side of Eq. \eqref{moments}. The relation in Eq. \eqref{conc0} 
has important conceptual consequences on which we will elaborate below.

\subsection{Brownian motion: Single-trajectory PSD}
\label{ST}

One infers from Fig. \ref{FIG1} that upon an  increase of  $f T$ the
oscillatory terms in $\gamma_S$ fade out,  and $\gamma_S$ saturates at
the value $\gamma_S = \sqrt{5}/2$.  Let us define $\omega_l$ as the
value of $f T$ when the amplitude of the oscillatory terms in $\gamma_S$
equals $\sqrt{5}/2 + \varepsilon$, where $\varepsilon > 0$ is a small fixed
number. Given that the amplitude of the oscillatory terms is a monotonically
decreasing function of $f T$, one has that for $f T > \omega_l$ the variation
coefficient $\gamma_S(f T > \omega_l) < \sqrt{5}/2 + \varepsilon$.  Next, the
decay law of the amplitude of oscillations can be readily derived from Eqs.
\eqref{power_av} and \eqref{var} to give that, in the leading in $\varepsilon$
order, $\omega_l \sim 2/(5 \varepsilon)$.

Then, for $f T > \omega_l$, up to terms proportional to $\varepsilon$, which can
be made arbitrarily small, we see that the moments of the random amplitude $A^{
(k)}(\gamma_S)$ in Eq.~\eqref{conc0} are given by
\begin{eqnarray}
\label{moments2}
\mathbb{E}\left\{\left(A^{(k)}(\gamma_S)\right)^Q\right\}&=&\frac{3^{Q+k/2}
\Gamma\left(Q + k\right)}{4^{Q+k/2}\Gamma\left(k\right)}\nonumber\\
&&\hspace*{-2.2cm}\times \,_2F_1\left(\frac{Q+k}{2},\frac{Q+k+1}{2};\frac{k+1}{2};
\frac{1}{4}\right),
\end{eqnarray}
meaning that in this limit $A^{(k)}(\gamma_S)$ becomes just a real number $A^{(k)}$ that is
independent of $f$ and $T$.
This implies, in turn, that 
with any necessary accuracy, prescribed by the choice of $\varepsilon$, and for any realization of a trajectory ${\bf X}_t$, 
\begin{align}
\label{concept}
\tilde{S}_T^{(k)}(f) =   \frac{4 D}{f^2}  A^{(k)} \left(1 + O\left(\varepsilon\right)\right)\,, \,\,\, \text{for $f T \in (\omega_l,\infty)$} \,,
\end{align}
where 
 the symbol $O(\varepsilon)$ signifies that the omitted terms, (stemming out of the oscillatory terms in $\mu_S(f, T)$, $\sigma^2_S(f,T)$ and hence, in $\gamma_S(f,T)$), 
have an amplitude smaller than $\varepsilon$.

Therefore, 
we arrive at the following conclusion, which is the main conceptual result of our analysis:
for continuous-time 
BM and $f T \in (\omega_l, \infty)$, 
the frequency-dependence of the 
PSD of any single trajectory is 
the same as of the ensemble-averaged PSD with any desired accuracy, set by $\varepsilon$.  
Consequently, in response to the title question of our
work, we conclude that for BM the $1/f^2$-dependence of the
power spectrum  can be deduced  
already from a single, sufficiently  long trajectory, without any necessity to perform an additional 
averaging over an ensemble of such trajectories. In Sec. \ref{discrete} (see Fig. \ref{FIG11}) below we 
show that this prediction made for BM---a somewhat idealized 
stochastic process with infinitesimal increments---holds indeed for discrete-time random walks and single-trajectory experiments, in which a
BM trajectory is recorded at discrete instants of time and for a finite
observation time.

Lastly, we note that since $\mu_S(f,\infty)$ is linearly proportional to the diffusion coefficient $D$  and the latter appears to be multiplied by a random numerical amplitude,  
one cannot infer the 
ensemble-averaged diffusion coefficient from a single-trajectory PSD. The error in estimating $D$ from a single trajectory is given precisely 
by the deviation of this amplitude from its averaged value. Below we discuss the fluctuations of this numerical amplitude.

\subsection{Fluctuations of the amplitude $A^{(k)}$}
\label{FL}

\begin{figure}
\includegraphics[width=8.7cm]{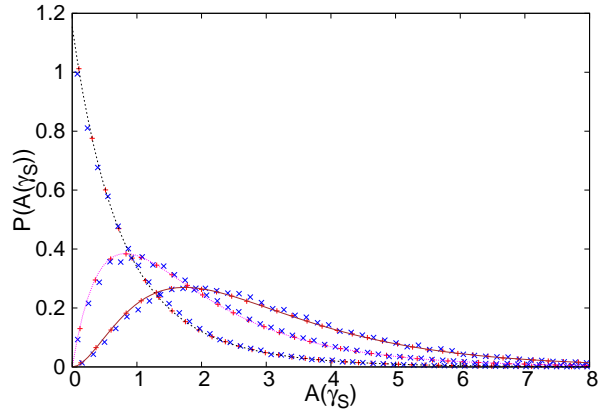}
\caption{ 
Distribution of the random amplitude $A^{(k)}(\gamma_S)$, defined in
Eq.~\eqref{conc0}. The curves depict the limiting distribution $P(A^{(k)}=A)$
in Eq.~\eqref{limiting} for $k=1$ (dashed), $k=2$ (dotted) and $k = 3$ (solid).
Symbols denote the results obtained for a BM generated by the truncated Wiener's
representation (crosses $\times$) with $N = 3.2 \times 10^6$ (see Sec.~\ref{TR})
and by Monte Carlo simulations (pluses $+$) of a discrete-time random walk (see
Sec.~\ref{discrete}) with  $T =1$ and $M = 2^{22}$ steps.
\label{FIG3}}
\end{figure}

The exact limiting 
probability density function of the amplitude $A^{(k)}$ in Eq. \eqref{concept} follows from Eq. \eqref{dist} and reads
\begin{align}
\label{limiting}
P(A^{(k)}=A) = \frac{2 \sqrt{\pi} \, A^{\frac{k-1}{2}}}{\sqrt{3} \, \Gamma\left(k/2\right)} 
 \exp\left(- \frac{4}{3} A\right) {\rm I}_{\frac{k-1}{2}}\left(\frac{2}{3} A\right) \,.
\end{align}
In Fig.~\ref{FIG3} we depict this distribution for $k=1,2$ and $3$ (curves from
left to right).  We observe that the very shape of the distribution depends
on the number of components which are taken into account in order to evaluate
the partial PSD. For $k=1$ (that is, for BM in one dimensional systems, or in
two or three-dimensions but when only one of the components is being tracked),
the distribution is a monotonically decreasing function with the maximal
value at $A^{(1)}=0$.  In contrast, for $k=2$ and $k=3$,  $P(A^{(k)})$ is a
bell-shaped function with a left power-law tail and an exponential right tail.

In Fig.~\ref{FIG3} we also present a comparison of our analytical predictions
in Eq.~\eqref{limiting} against the results of numerical simulations. We use
two methods to produce numerically the distributions of $A^{(k)}(\gamma_S)$
defined in Eq.~\eqref{conc0} for the range of frequencies where $f^2
S_T(f)$ is almost constant (see Secs. \ref{TR} and \ref{discrete} for more
details). The first method hinges on Wiener's representation of a BM in
Eq.~\eqref{wiener}, which is truncated at the upper summation limit at $N =
3.2 \times 10^6$ (see Sec. \ref{TR} below for more details). Crosses ($\times$)
in Fig. \ref{FIG3} depict the corresponding results obtained via averaging over
$10^5$ trajectories generated using such a representation of BM.  Further,
we use a discrete-time representation of a BM---lattice random walks with
unit spacing and stepping events at each tick of the clock---which are produced by
Monte Carlos simulations.  We set the observation time $T = 1$ and generate
trajectories with $M = 2^{22}$ steps to get a single-trajectory periodogram. A
thorough discussion of the domain of frequencies in which the periodogram
yields the behavior specific to a continuous-time BM are presented below in
Sec.~\ref{discrete}. Pluses ($+$) depict the results obtained numerically for
random walks averaged over $10^5$ realizations of the process.  Overall,
we observe an excellent agreement between our analytical predictions,
derived for BM---a continuous-time process with infinitesimal increments,
and the numerical results. This signifies that the framework developed here
is completely justified and can be used for a meaningful interpretation of
stochastic trajectories obtained in single-trajectory experiments.

Moreover, the moments of the distribution in Eq.~\eqref{limiting} are given by
Eq.~\eqref{moments2}. We find that the average value is $\mathbb{E}\{A^{(k)}\}
= k$, as it should be, and the variance ${\rm Var}\left(A^{(k)}\right) = 5
k/4$ so that the distribution broadens with increasing $k$. The coefficient
of variation of the $k$-dependent distribution in Eq.~\eqref{limiting} is equal
to  $\sqrt{5/(4 k)}$ meaning that fluctuations become progressively less
important for increasing $k$. 
We also remark that
the most probable values for the cases $k=2$ and $k=3$ are $A_{mp}^{(2)}
= 3 \ln(3)/4 \approx 0.82$ and $A_{mp}^{(3)} \approx 1.74$, respectively,
and are well below their average values.

\begin{figure}
\includegraphics[width=8.5cm]{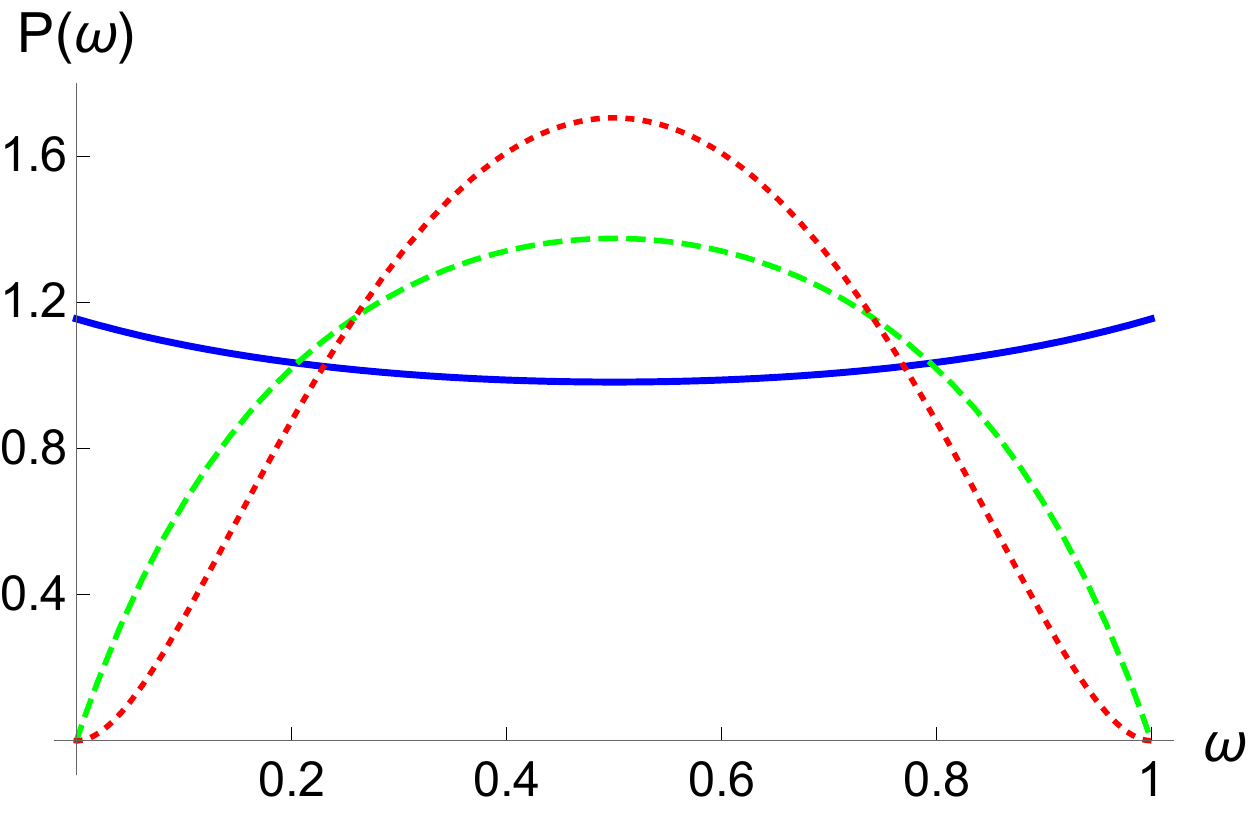}
\caption{Probability density function $P(\omega^{(k)}=\omega)$
for $k=1$ (solid line), $k=2$ (dashed line) and $k=3$ (dotted line). 
\label{F}}
\end{figure} 

Lastly, we quantify the effective broadness of the distribution in Eq. \eqref{limiting},
focusing  on its heterogeneity index 
\begin{align}
\omega^{(k)} = \frac{ A_1^{(k)}}{A_1^{(k)}+A_2^{(k)}},
\end{align}
where $A_1^{(k)}$ and $A_2^{(k)}$ are the amplitudes  drawn from two different independent realizations
 of  ${\bf X}_t$. Such
a diagnostic tool has been proposed in Refs. \cite{90,91,92} 
in order to
quantify fluctuations in the first passage phenomena in bounded domains, compare
also with the discussion in Ref.~\cite{aljaz,aljaz1}. 
In our context, $\omega^{(k)}$  shows the likeliness of the event that two values of the amplitudes $A_1^{(k)}$ and $A_2^{(k)}$ will be equal to each other.  
With the distribution \eqref{limiting} of the random variable $A^{(k)}$  the distribution $P(\omega^{(k)})$ of the heterogeneity index
can be calculated via its integral representation  \cite{90,91}
\begin{align}
\label{k}
P(\omega^{(k)} = \omega) &= \frac{1}{(1 - \omega)^2} \int^{\infty}_0 A^{(k)} dA^{(k)} \, \nonumber\\
& \times P\left(A^{(k)}\right) \, P\left(\frac{\omega}{1 - \omega} A^{(k)}\right) \,.
\end{align}
Performing the integrals in relation \eqref{k}, we find the following exact expressions for the distribution of the heterogeneity index: For $k=1$ we have
\begin{align}
\label{158}
&P\left(\omega^{(1)} = \omega\right)
= \frac{2 \sqrt{3}}{\pi \sqrt{\omega \left(3 + \omega \left(1 - 4 \omega(2 - \omega)\right)\right)}} \nonumber\\
& \times \frac{{\bf E}\left(\exp(- \eta)\right)}{\sinh\left(\eta/2\right)} \,,  \,\,\,
\eta = \operatorname{arcosh}\left(1+\frac{3}{2 \omega (1-\omega)}\right) \,,
\end{align}
where ${\bf E}\left(\ldots\right)$ is the complete elliptic integral of the second kind; for $k=2$ the distribution has the form
\begin{align}
\label{168}
P\left(\omega^{(2)} = \omega\right) = \frac{2 \omega \left(1 - \omega\right) \left(39 - 20 \omega (1 -  \omega)\right)}{\left(3 - 2 \omega \right)^2 \left(1 + 2 \omega\right)^2} \,,
\end{align}
while for $k=3$ it follows that
\begin{align}
\label{3D}
P\left(\omega^{(3)} = \omega\right) &= - \frac{27}{16} \sqrt{\omega (1 - \omega) }  \Bigg[\frac{d^3}{dp^3} \exp\left( - \frac{3}{2} \eta^*\right) \nonumber\\
&\times  _2F_1\left(\frac{1}{2}, \frac{3}{2}; 2 ; \exp\left(- 2 \eta^*\right)\right)\Bigg]_{p=1} \,,
\end{align}
with 
\begin{align}
\eta^* = \operatorname{arcosh}\left(\frac{4 p^2 - (1 - \omega)^2 - \omega^2}{2 \omega (1 - \omega)}\right) \,.
\end{align}
Note that the third-order derivative with respect to $p$ in relation \eqref{3D}
can be taken explicitly producing, however, a rather cumbersome expression 
in terms of the complete elliptic integrals. For the sake of compactness, we
nonetheless prefer the current notation.

The probability density functions in Eqs. \eqref{158}, \eqref{168} and
\eqref{3D} are depicted in Fig. \ref{F}. We observe that for
$k=1$ the event in which  two values of $A^{(1)}$ deduced from two different
independent trajectories are equal to each other, (that is, when $A_1^{(1)} =
A_2^{(1)}$ and  $\omega = 1/2$), is the  most unlikely, since it corresponds
to the minimum of the distribution $P\left(\omega^{(1)} = \omega\right)$
in Eq. \eqref{158}. Therefore, for the case $k=1$ the most probable outcome
is that two values of $A^{(1)}$ obtained for two different realizations of
BM will be very different from each other. For $k=2$ and $k=3$, $\omega =
1/2$ corresponds to the most probable event but still the distributions in
Eqs. \eqref{168} and \eqref{3D} appear to be rather broad so that a pronounced realization-to-realization
variation of the amplitude is expected.

\section{Truncated Wiener's representation}
\label{TR}

In simulations of a continuous-time BM one often uses Wiener's representation
\eqref{wiener}, truncating it at the upper summation limit at some integer $N$,   
\begin{align}
\label{wiener1}
X^{(tr)}_t = \frac{\sqrt{2 T}}{\pi} \sum_{n=1}^{N} \frac{\zeta_n}{(n - 1/2)} \sin\left(\frac{(n - 1/2) \pi t}{T}\right) \,.
\end{align}
The sample paths of such a partial sum process are known to converge
to the sample paths of the BM at the rate $1/\sqrt{N}$.  Focusing on a
single-component single-trajectory PSD in Eq.  \eqref{single} (generalization
to the $d$-dimensional case and a partial PSD $\tilde{S}^{(k)}_T(f)$ in
Eq. \eqref{partial} is straightforward) we have  the following estimate for
$S^{(j)}_{T}(f) $,
\begin{align}
\label{def21}
&S^{(tr)}_{T}(f)=\frac{2}{\pi^2}\left[\left(\sum_{n=1}^Ng_n\zeta_n\right)^2+
\left(\sum_{n=1}^{N} h_n \zeta_n\right)^2 \right]. 
\end{align}
Below we examine how accurately $S^{(tr)}_{T}(f)$ reproduces $S^{(j)}_{T}(f)$. We
first consider the first moment of the truncated series and its variance, which
are given by
\begin{subequations}
\begin{align}
\label{trunc}
\mu_{tr}(f,T) = \frac{4 D}{\pi^2} \sum_{n=1}^N \left(g_n^2 + h_n^2\right)
\end{align}
and
\begin{align}
\sigma^2_{tr}(f,T) &= \frac{32 D^2}{\pi^4} \Bigg[\left(\sum_{n=1}^N g_n^2\right)^2 + \left(\sum_{n=1}^N h_n^2\right)^2+ \nonumber\\
&+ 2 \left(\sum_{n=1}^N g_n h_n\right)^2\Bigg] \,.
\end{align}
\end{subequations}

\begin{figure}
\includegraphics[width=8.5cm]{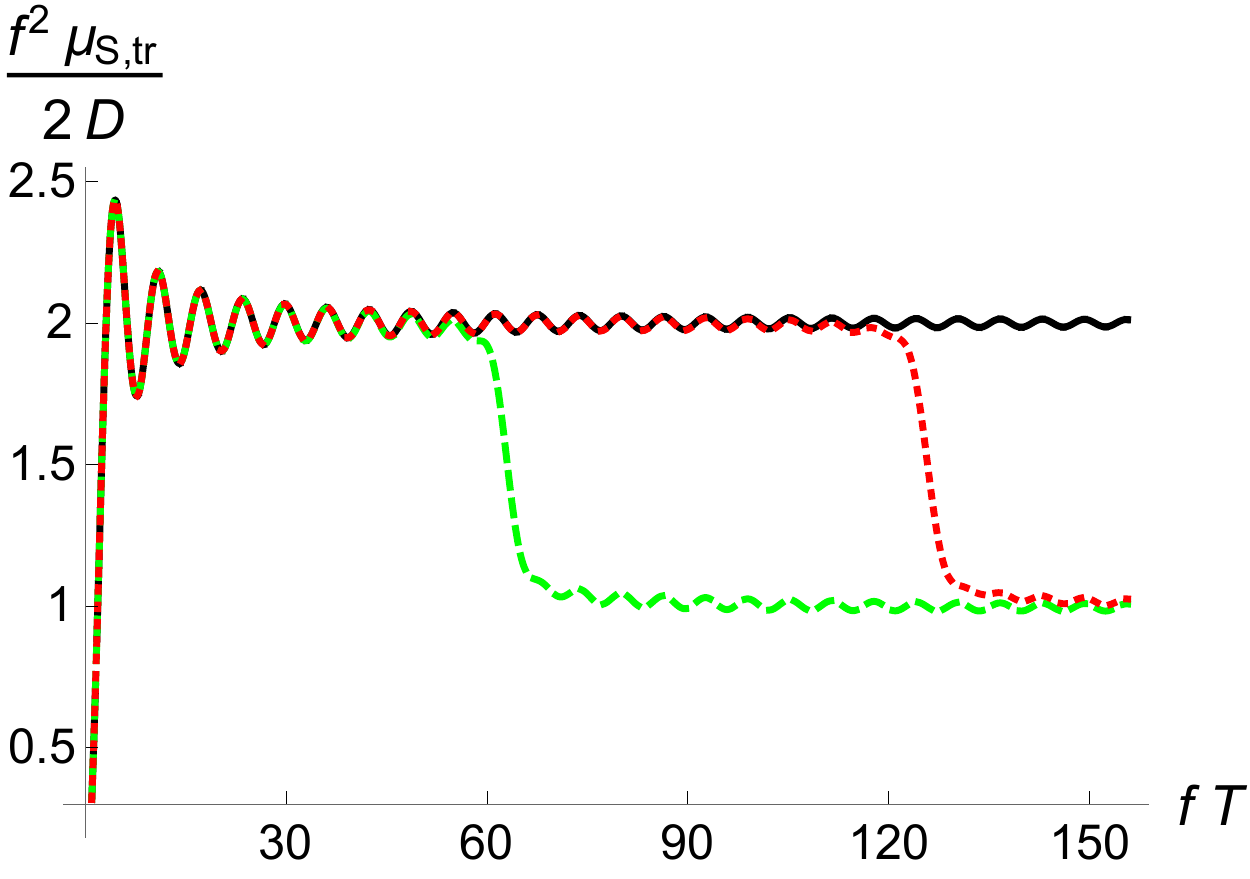}
\includegraphics[width=8.5cm]{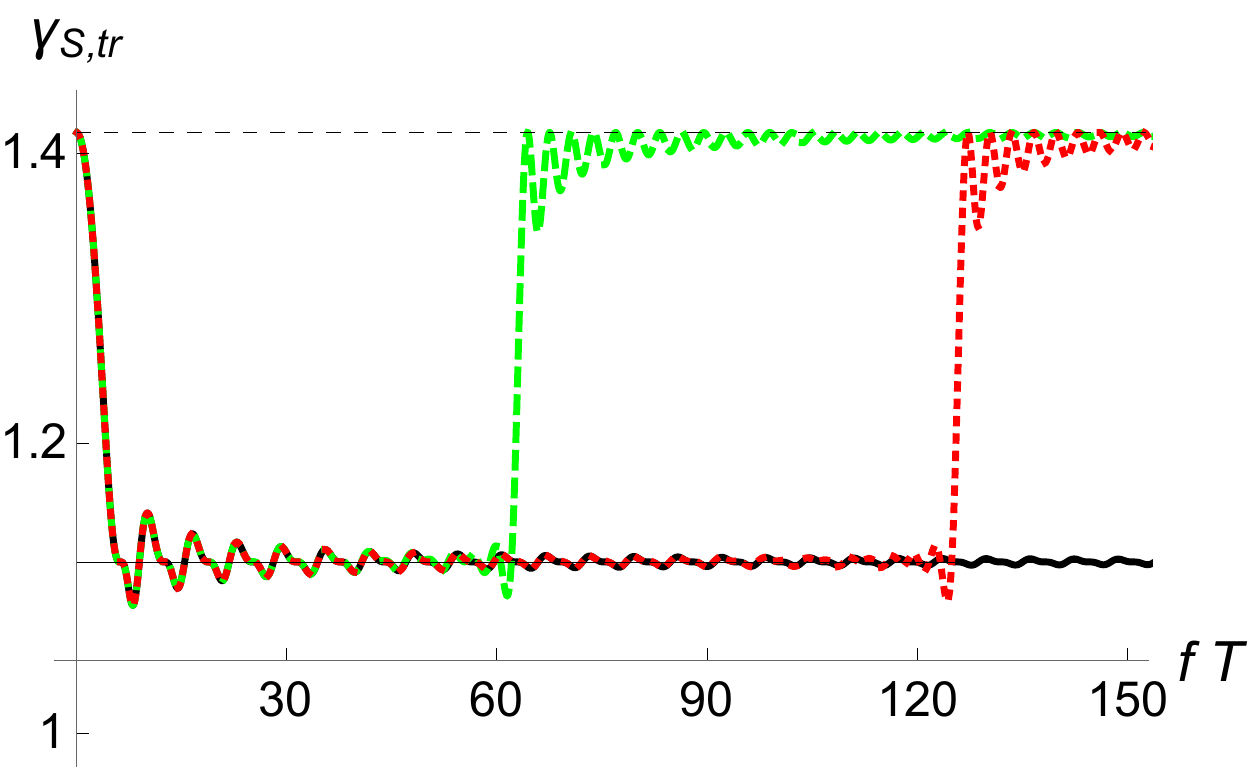}
\caption{Upper panel: Comparison of the first moment
$\mu_S(f,T)$ of a single-component single-trajectory PSD  and its analog
$\mu_{tr}(f,T)$ in Eq.~\eqref{trunc}, obtained by truncating the series
in Wiener's representation at an integer $N$, as functions of $f
T$. The solid curve represents $\mu_S(f,T)$. The dashed (green) curve
corresponds to $\mu_{tr}(f,T)$ with $N=20$, the dotted (red) curve to
$N=40$. Lower panel: Analogous results for the coefficients of variation
$\gamma_S = \sigma_S(f,T)/\mu_S(f,T)$ (solid curve) and $\gamma_{tr} =
\sigma_{tr}(f,T)/\mu_{tr}(f,T)$ (dashed, $N=20$ and dotted, $N = 40$,
curves). Thin horizontal lines correspond to $\sqrt{5}/2$ (solid) and
$\sqrt{2}$ (dashed).
\label{FIG30}}
\end{figure}

In Fig. \ref{FIG30} we present a comparison of the averaged single-component
single-trajectory PSD and of its counterpart obtained from the truncated
Wiener's series \eqref{wiener1}, as well as of the corresponding coefficients
of variation of the two probability density functions.  We observe a perfect
agreement between the results obtained from the complete series and the
truncated ones. Moreover, we see that $\mu_{tr}(f,T)$ and $\gamma_{tr}$
exhibit a \textit{uniform} convergence to $\mu_S(f,T)$ and $\gamma_S$,
respectively, for $f T \in (0, \pi N)$. This implies that keeping just $N =
40$ terms in the truncated Wiener's series permits us to describe reliably
well the behavior of the pertinent properties over more than two decades
of variation of  $f T$. Extending this interval up to three decades will
require keeping $N \approx 320$ terms, for four decades $N \approx 3200$
terms, and so on.

We finally focus on the moment-generating function $\Phi_{\lambda}(S^{(tr)}_{
T}(f))$ of  $S^{(tr)}_{T}(f)$, and find
\begin{align}
\label{MGtrun}
\Phi_{\lambda} &= \Bigg[1 + 2  \, \mu_{tr}(f,T) \, \lambda +  \left(2  - \gamma^2_{tr}\right) \, \mu^2_{tr}(f,T) \,\lambda^2\Bigg]^{-1/2} \,.
\end{align}  
We note that $\Phi_{\lambda}(S^{(tr)}_{T}(f))$ has exactly the same form as the moment-generating function \eqref{MG} (with $k=1$)
evaluated for a complete Wiener's series and hence, $S^{(tr)}_{T}(f)$ has the distribution of exactly the same form as the one in Eq. \eqref{dist}
with the only difference that 
the first moment and the variance 
have to be replaced by their counterparts obtained via truncation of the Wiener's representation at some level $N$. 
Given that for $f T \in (0,\pi N)$ these properties are identical (see Fig. \ref{FIG30}), it means that in this interval of variation 
of $f T$ equation  
\eqref{MGtrun}
coincides with \eqref{MG}, and the probability density function of $S^{(tr)}_{T}(f)$ coincides with result 
\eqref{dist}. Outside of this interval, i.e., for $f T > \pi N$, 
the coefficient of variation of the truncated PSD jumps from $\sqrt{5}/2$ to $\sqrt{2}$ meaning that the distribution of $S^{(tr)}_{T}(f)$ becomes the $\chi^2$-distribution (see Appendix \ref{A}), which
is evidently a spurious behavior.

\section{Discrete sets of data}
\label{discrete}

In experiments or in Monte Carlo simulations of BM, the particle position is
recorded  at some \emph{discrete\/} time moments such that one stores a given
trajectory as a finite set of data. Here we analyze how the features unveiled
in the previous Sections will change, if instead of continuous-time BM we rather 
use a picture based on a discrete-time random walk.
 
Suppose that the time interval $(0,T)$ is divided into $M$ equally-sized
subintervals $\Delta = T/M$, such that the particle position is recorded at
time moments $t_m = \Delta m$, $m =1, 2, \ldots, M$. We use the convention
that at $t=0$ the particle starts at the origin and focus on the behavior of a
single-component PSD---the extension of our analysis over the general case of
$k$ components is straightforward but results in rather cumbersome expressions.

As a first step, we convert the integrals in Eq.~\eqref{single} into the
corresponding sums to get a \emph{periodogram}
\begin{align}
\label{2}
R_M(f) = \frac{\triangle^2}{T} \sum_{m_{1,2}=0}^{M} \cos\left(f \triangle \left(m_1 - m_2\right)\right) \, X_{\triangle m_1} X_{\triangle m_2} \,,
\end{align}
where we now denote a single-trajectory 
PSD as $R_M(f)$ to emphasize that it is  a different mathematical object as compared to the PSD in Eq. \eqref{single}. 
Since we now have only a finite amount of points instead of a continuum, $R_T(f)$ in Eq. \eqref{2} will be a periodic function of $f T$ so that
it may 
only  
\textit{approximate} the behavior of the 
PSD for continuous-time BM in some range of frequencies at a given observation time.
Further on, we use the scaling property of BM to rewrite the latter expression as
\begin{align}
\label{3}
R_M(f) = \frac{2 D \triangle^3}{T} \sum_{m_{1,2}=0}^{M} \cos\left(f \triangle \left(m_1 - m_2\right)\right) \, B_{m_1} B_{m_2} \,,
\end{align}
where $B_m$ is now a trajectory of a standard lattice random walk
with unit spacing and stepping at each clock tick $m$, $m = 0,1,2, \ldots, M $.
Next, we write
\begin{align}
B_m = \sum_{j = 0}^m s_j,
\end{align}
where $s_j=\pm1$ are independent increments, and $s_0 \equiv 0$. Then, expression
\eqref{3} becomes
\begin{align}
\label{disc}
R_M(f) &= 
\frac{2 D \Delta^2}{M} \Bigg[\left(\sum_{j=1}^M  a_j s_j\right)^2 + \left(\sum_{j=1}^M b_j s_j\right)^2\Bigg] \,,
\end{align}
where
\setcounter{equation}{41}
\begin{subequations}
\begin{eqnarray}
\label{aj}
a_j&=&\sum_{m=j}^M\cos\left(f\Delta m\right)=\frac{1}{2} \Bigg(\cos\left(f\Delta
j\right)+\cos\left(f\Delta M\right)\nonumber\\
&&+\cot\left(f\Delta/2\right)\Big(\sin\left(f\Delta M\right)-\sin\left(f\Delta
j\right)\Big)\Bigg)
\end{eqnarray}
and
\begin{eqnarray}
\label{bj}
b_j&=&\sum_{m=j}^M \sin\left(f\Delta m\right)=\frac{1}{2} \Bigg(\sin\left(f\Delta
j\right)+\sin\left(f\Delta M\right)\nonumber\\
&&\hspace*{-0.4cm}
+\cot\left(f\Delta/2\right)\Big(\cos\left(f\Delta j\right)-\cos\left(f\Delta
M\right)\Big)\Bigg).
\end{eqnarray}
\end{subequations}
Expression \eqref{disc} is the discrete-time analog of the PSD in Eq.~\eqref{def2}
obtained for the continuous-time BM.

\subsection{Discrete-time case: Mean and variance of a single-trajectory
periodogram}
\label{Z}  

At this point, it may be expedient to first look at the ensemble-averaged
single-trajectory periodogram in Eq. \eqref{disc} and at its variance, and
to compare them against their continuous-time counterparts. Averaging the
expression in Eq.~\eqref{disc} and its squared value over all possible
realizations of the increments $s_j$, we have
\begin{widetext}
\begin{align}
\label{dis_mean}
&\mu_R(f,T) = \frac{2 D \Delta^2}{M} \sum_{j=1}^M \left(a_j^2 + b_j^2\right) =
\frac{D \Delta^2}{\sin^2\left(f \Delta/2\right)}  \Bigg[1 + \frac{1}{2 \, M} - \frac{\Big(\sin\left(f \Delta M\right)+ \sin\left(f \Delta (M + 1)\right)\Big)}{2 \, M \,  \sin\left(f \Delta\right)}\Bigg] \,,
\end{align}
and
\begin{align}
\label{dis_var}
&\sigma^2_R(f,T)  = \frac{8 D^2 \Delta^4}{M^2} \Bigg[\left(\sum_{j=1}^M a_j^2\right)^2 - \sum_{j=1}^M a_j^4
+\left(\sum_{j=1}^M b_j^2\right)^2 - \sum_{j=1}^M b_j^4 + 2  \left(\sum_{j=1}^M a_j \, b_j\right)^2 - 
2 \sum_{j=1}^M a_j^2 \, b_j^2 
\Bigg] = \nonumber\\
&= \frac{20 D^2 \Delta^4 \cos^4\left(f \Delta/2\right)}{\sin^4\left(f \Delta\right) } \Bigg[1 
- \frac{1}{10 M \sin^2\left(f \Delta\right)} \Big(7 - 7 \cos\left(2 f \Delta\right) + 6 \cos\left(f \Delta M\right) - \cos\left(2 f \Delta M\right) + \cos\left(2 f \Delta \left(M + 1\right)\right) 
\nonumber\\
& - 6 \cos\left(f \Delta \left(M + 2\right)\right) + 12 \sin\left(f \Delta\right) \sin\left(f \Delta M\right)\Big) + 
 \frac{1}{10 M^2 \sin^2\left(f \Delta\right)} \Big(4 + 5 \cos\left(2 f \Delta\right) + 
 8 \cos\left(f \Delta\right) \left(1 - 
\cos\left(f \Delta M\right)\right) \nonumber\\ 
&- 2 \cos\left(f \Delta  M\right)  - 2 \cos\left(2 f \Delta M\right) + \cos\left(2 f \Delta \left(M+1\right)\right) - 6 \cos\left(f \Delta \left(M + 2\right)\right) + 16 
\sin\left(f \Delta\right) \sin\left(f \Delta M\right)\Big) 
\Bigg] \,.
\end{align}
\end{widetext}
Note that these somewhat lengthy expressions \eqref{dis_mean} and \eqref{dis_var} are exact, and valid for any $M$, $\Delta$ and $f$. 

\begin{figure}
\includegraphics[width=8.5cm]{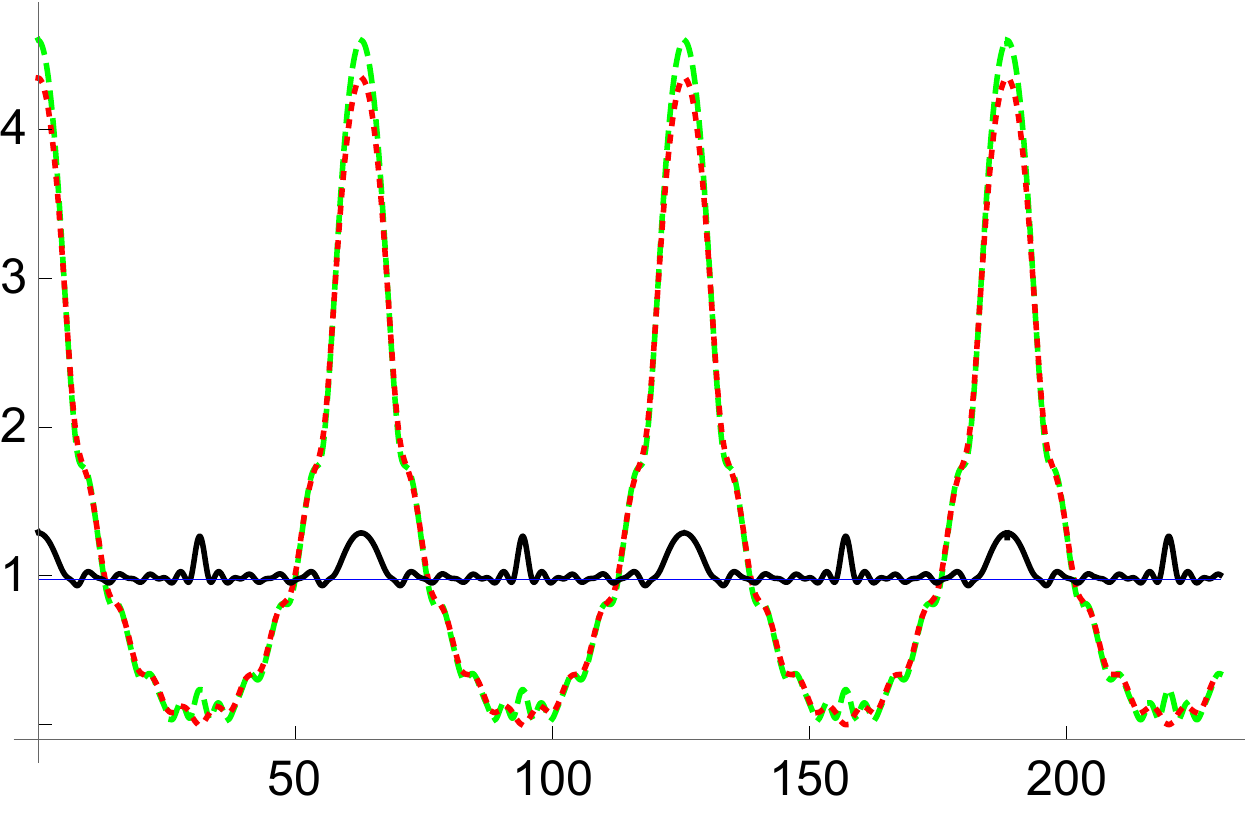}
\caption{Plot of $\ln\left(\mu_R(f,T)/(D \Delta^2)\right)$  in Eq. \eqref{dis_mean} (red dotted curve), $\ln\left(\sigma_R(f,T)/(D \Delta^2)\right)$  in Eq. \eqref{dis_var} (green dashed curve) and $\gamma_R = \sigma_R(f,T)/\mu_R(f,T)$ (solid curve) as functions of  $f T$ for $M = 10$. The thin horizontal line is $\gamma_R = \sqrt{5} \sqrt{1-12/(5M)}/2$, see Eq. \eqref{zu}. Note that we plot the logarithms of $\mu_R(f,T)$ and $\sigma_R(f,T)$ instead of these properties themselves in order to be able to show their full variation along the ordinate axis.
\label{sketch}
}
\end{figure} 

To illustrate the behavior of $\mu_R(f,T)$, $\sigma_R(f,T)$ and of the
corresponding coefficient of variation $\gamma_R = \sigma_R(f,T)/\mu_R(f,T)$ of
the distribution of a single-trajectory periodogram $R_M(f)$, we first depict
in Fig.~\ref{sketch} these parameters as functions of $f T$ for a rather small 
value of $M$, $M=10$. We observe that they are periodic functions of $f T$ with
the prime period $2\pi M$, (such that Fig.~\ref{FIG1}  presents a zoom of just
one-half of the prime period) and exhibit \emph{peaks\/} in the center, that is,
at $f T=\pi M$, and at the end-points of the prime period, that is, for $fT=0$
and $fT=2\pi M$. More specifically, we see that in the middle of the prime period
\begin{subequations}
\begin{align}
\label{peaks1}
\mu_R\left(\frac{\pi M}{T},T\right)=D\Delta^2 
\end{align}
and
\begin{align}
\label{peaks2}
\sigma^2_R\left(\frac{\pi M}{T},T\right) = 2 D^2 \Delta^4 \left(1 - \frac{2}{M}\right)  \,,
 \end{align}
leading to
\begin{align}
\label{peaks3}
\gamma_R = \sqrt{2} \sqrt{1 - \frac{2}{M}} \,.
\end{align}
\end{subequations}
For fixed $T$ and $M \to \infty$,   $\mu_R\left(\pi M/T,T\right)$ and $\sigma_R\left(\pi M/T,T\right)$ both tend to zero, 
but their ratio $\gamma_R$  stays finite and
tends to  $\sqrt{2}$.
Next, at the end-points of the prime period 
 $\mu_R(f,T)$ and $\sigma_R(f,T)$ are given by 
\begin{subequations}
\begin{align}
\label{atperiod}
\mu_R\left(\frac{2 \pi M}{T},T\right) = \frac{2 D T^2}{3} \left(1 + \frac{3}{2 M} + \frac{1}{2 M^2}\right)
\end{align} 
and
\begin{align}
\label{atperiod1}
\sigma_R^2\left(\frac{2 \pi M}{T},T\right) &= \frac{8 D^2 T^4}{9} \nonumber\\
&\times  \left(1 + \frac{6}{5 M} - \frac{5}{4 M^2} + O\left(\frac{1}{M^3}\right)\right) \,,
\end{align}
such that they both depend on the observation time $T$ and diverge when $T \to \infty$. In this case
 $\gamma_R$ obeys
\begin{align}
\label{atperiod2}
\gamma_R = \sqrt{2} \left(1 - \frac{9}{10 M} + \frac{9}{200 M^2}  + O\left(\frac{1}{M^3}\right)\right) \,,
\end{align}
and also stays finite as $M \to \infty$.
\end{subequations}

Consider next the behavior of $\mu_R(f,T)$ and $\sigma_R(f,T)$ for $fT=\pi M/2$,
that is, at one-quarter of the prime period, and also for $fT=2\pi M/3$, that is,
at one-third of the prime period. Here, we have
\begin{subequations}
\begin{align}
\label{14}
\mu_R\left(\frac{\pi M}{2 T},T\right) = 2 D \Delta^2
\end{align} 
and
\begin{align}
&\sigma_R^2\left(\frac{\pi M}{2 T},T\right)=5D^2\Delta^4\left(1-\frac{12}{5M}
\right),
\end{align}
producing
\begin{align}
\label{zu}
\gamma_R = \frac{\sqrt{5}}{2} \sqrt{1 - \frac{12}{5 M}}  \,,
\end{align}
\end{subequations}
and
\begin{subequations}
\begin{align}
\label{onethird1}
\mu_R\left(\frac{2 \pi M}{3 T},T\right) = \frac{4 D \Delta^2}{3} \,,
\end{align} 
\begin{align}
\label{onethird2}
&\sigma_R^2\left(\frac{2 \pi M}{3 T},T\right) = \frac{20 D^2 \Delta^4}{9} \left(1 - \frac{12}{5 M} \right)  \,,
\end{align}
\end{subequations}
which imply that here, as well, $\gamma_R$ is given by Eq.~(\ref{zu}).
Therefore, we have that for the last two situations the first moment and the
variance of a single-trajectory periodogram both tend to zero when $\Delta
\to 0$ (that is, when $M \to \infty$ at a fixed $T$), which is similar to
the behavior at one-half of the prime period. In contrast, here  $\gamma_R$
tends to $\sqrt{5}/2$, which is the asymptotic coefficient of variation of
the distribution of a single-trajectory PSD for the continuous-time BM.

\begin{figure}
\includegraphics[width=8.5cm]{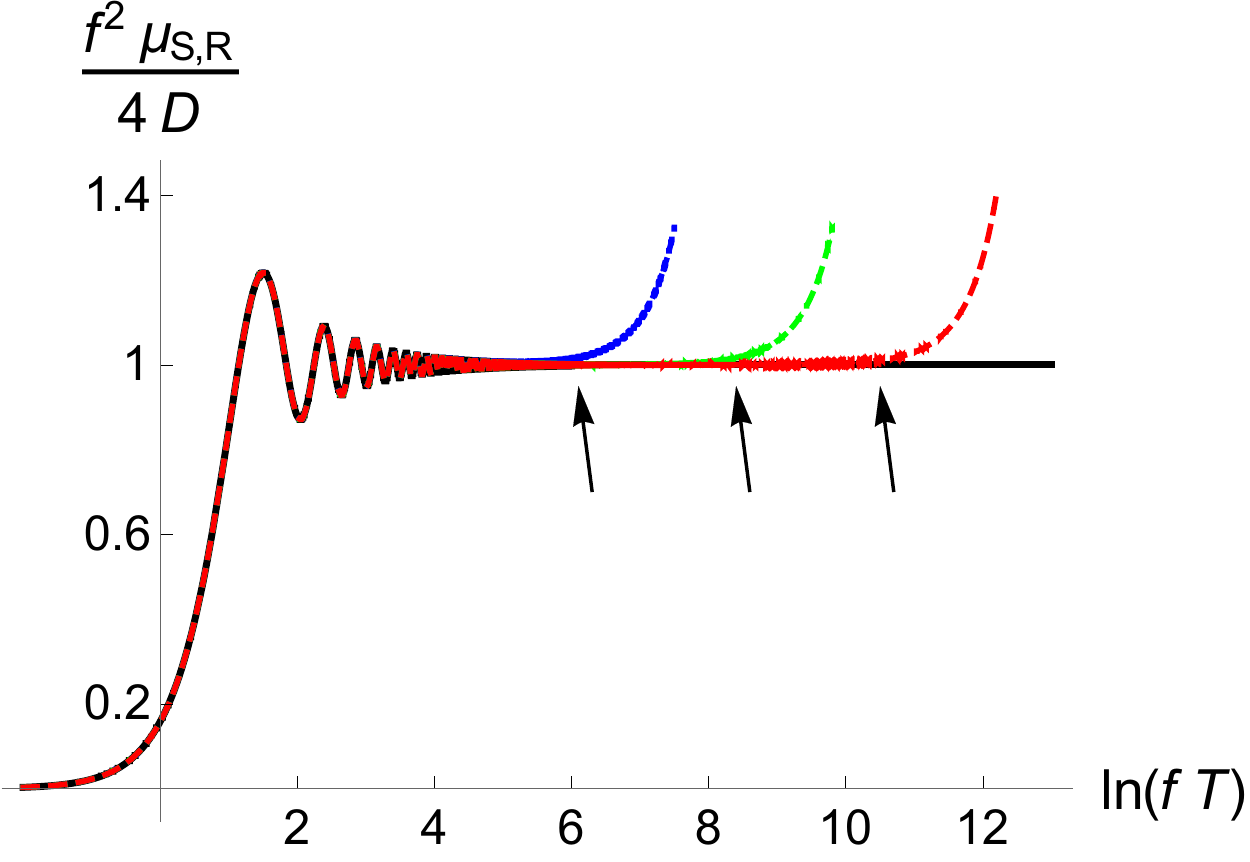}
\caption{ 
Comparison of the averaged PSD in Eq. \eqref{power_av} for the continuous-time BM  (solid curve) and for the discrete
case, Eq. \eqref{dis_mean}, (dashed curves), with $M = 10^3$ (blue), $M = 10^4$ (green) and $M = 10^5$ (red).
The reduced first moments $f^2 \mu_{S,R}(f,T)/4D$ of the PSD for the continuous- and the discrete-time 
cases are plotted versus the logarithm of  $f T$ over several decades of variation of this parameter.
 Arrows indicate the values of  $f T$ at which the periodicity of  $R_M(f)$ in Eq. \eqref{3}  starts to matter in a noticeable way.
\label{FIG7}
}
\end{figure}

In Fig. \ref{FIG7} we present a comparison between the averaged PSD
\eqref{power_av} for the continuous-time BM and its discrete-time counterpart
\eqref{dis_mean}. We observe a nearly perfect agreement between these two
representations for a wide range of variation of $f
T$. We see that the periodicity of  $\mu_R(f,T)$ starts to matter in a
noticeable way only at $f T \approx 4 \times 10^2$ for $M=10^3$, at $f T
\approx 3 \times 10^3$ for $M=10^4$ and only at $f T \approx 2 \times 10^4$
for $M=10^5$, respectively.  Below we will quantify the onset of the deviation
between Eqs. \eqref{dis_mean} and  \eqref{power_av} more accurately.

Next, we come back to Fig. \ref{FIG1} in which we compare the coefficient
$\gamma_S$ of variation of the distribution of a single-trajectory PSD for
the case of the continuous-time BM, and an analogous coefficient $\gamma_R$
of variation of the distribution in the discrete-time case, calculated
using our results \eqref{dis_mean}  and \eqref{dis_var} for $M=10^4$ and
$M=10^5$. Here we observe that the agreement is even better and extends
over a fairly  large range of values of $f T$, until $\gamma_R$ starts to
show a strong oscillatory behavior close to the middle of the prime period
(see also Fig. \ref{sketch}).  Overall, it appears that $\gamma_R$ can be set
equal to $\sqrt{5}/2$ with an arbitrary accuracy dependent on the  choice of
$\varepsilon$ (see Sec. \ref{ST}) on a bounded interval $f T \in (\omega_l,
\pi M - \omega_l)$ and, by symmetry, for  $f T \in (\pi M + \omega_l, 2 \pi
M - \omega_l)$.

Therefore, in response to the title question of this work we conclude that 
one can indeed observe the behavior specific to the continuous-time BM
(that is, the $1/f^2$ dependence of the averaged PSD and, by virtue of
Eq.~\eqref{concept}, of a single-trajectory PSD) in the discrete-time
settings. However, the spectrum must be analyzed on a finite interval
$(\omega_l,\omega_r)$ of variation of $f T$, which is bounded from below by
$\omega_l$ (see Sec. \ref{ST})---ensuring that $\gamma_R$ becomes equal (with
any necessary accuracy) to $\sqrt{5}/2$ and hence, does not contribute to the
frequency dependence of a single trajectory PSD---and by $\omega_r$ from above,
when $\mu_R(f,T)$ starts to deviate from $\mu_S(f,T)$. Extending the arguments
presented in Sec.~\ref{ST}, we define $\omega_r$ as the value of $fT$ at which
$\mu_R(f,T)/\mu_S(f,T)=1+\varepsilon$. To calculate $\omega_r$, we note that
the ratio of two ensemble-averaged PSDs can be very accurately approximated by
\begin{align}
\label{freq}
\frac{\mu_R(f,T)}{\mu_S(f,T)} \approx \frac{(f T)^2}{4 M^2 \sin^2\left(f T/2 M\right)} \,,
\end{align} 
so that we find that $\omega_r = 2 \alpha M$, where $\alpha $ obeys
$\alpha^2/\sin^2(\alpha) = 1 + \varepsilon$. For small $\varepsilon$,
the parameter $\alpha$ is given, to leading order in $\varepsilon$, by
$\alpha = \sqrt{3 \varepsilon}$. Recalling the definition of $\omega_l$
(see Sec. \ref{ST}), we thus see that the $1/f^2$-dependence can be
observed once $\omega_r/\omega_l \gg 1$, which means that the number $M$
of recorded positions of each trajectory has to obey $M \gg 1/(5 \sqrt{3}
\varepsilon^{3/2})$.

\begin{figure}
\includegraphics[width=8.8cm]{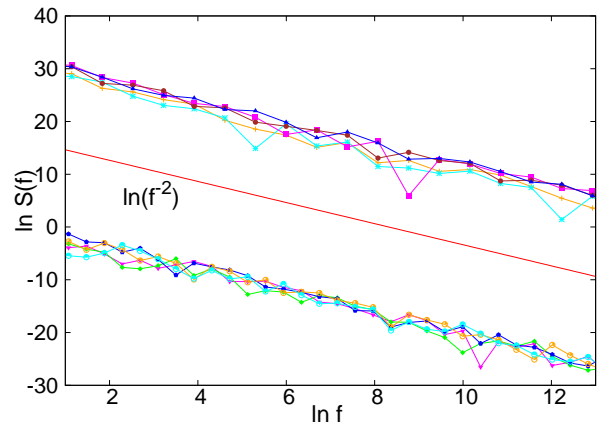}
\caption{ 
Logarithm of the power spectral density of five individual trajectories as
a function of $\ln(f T)$.  The upper curves are the numerical results based
on the Monte Carlo simulations, while the lower set of curves corresponds
to a BM generated using a truncated Wiener's representation.
\label{FIG11}}
\end{figure}

In Fig.~\ref{FIG11} we plot the logarithm of a single-trajectory periodogram
versus the logarithm of $fT$ for five individual, randomly chosen trajectories
of discrete-time lattice random walks and also five trajectories of a BM
generated by the truncated Wiener's series (see Sec.~\ref{TR}). For random
walks, (which make a step of unit length at each tick of the clock, so that
$D = 1/2$), we set the observation time $T = 1$ and have $M = 2^{22} \approx
4.2 \times 10^6$ discrete points within the unit time interval.  For such
a choice of parameters, we set the accuracy parameter $\varepsilon = 0.01$
to get $\omega_l = 40$ and $\omega_r = 1.45 \times 10^6$.  We observe that,
indeed, for all these five trajectories and $\ln f \in (\ln \omega_l = 3.69,
\ln \omega_r = 14.19)$, (that is, for $f$ varying over more than five decades),
the relation $\tilde{S}_T^{(k)}(f) \sim 1/f^2$ holds. Next, for a BM generated
by the truncated Wiener's representation we set $T = 1$, $D = 0.01$ and take
$N = 3.2 \times 10^6$. Here, as well, for all five trajectories we observe a
perfect agreement with our prediction in Eq. \eqref{concept} for $\ln f \in
(\ln\omega_l=3.69$, $\ln\pi N=14.98)$, that is, for $f$ spanning over more
than five decades.

\begin{figure}
\includegraphics[width=8.5cm]{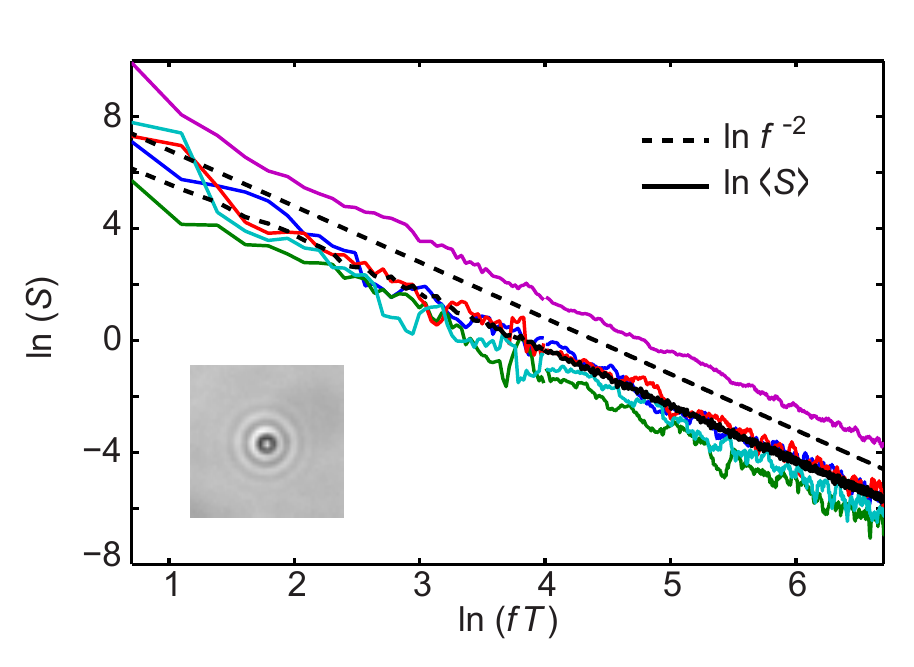}
\includegraphics[width=8.5cm]{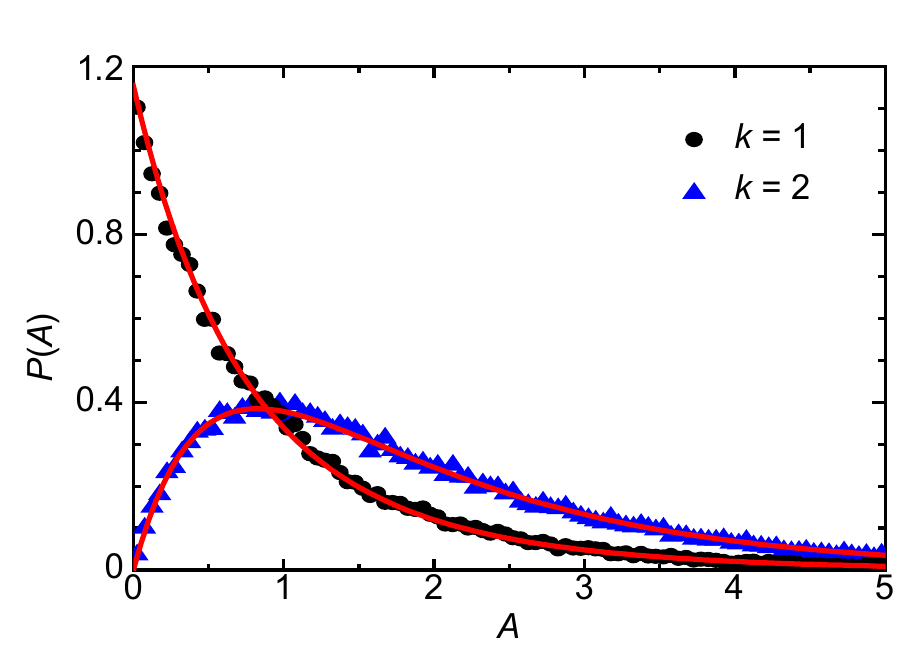}
\caption{ 
Brownian motion performed by 1.2-$\mu$m polystyrene beads in aqueous
solution. Upper panel: Logarithm of the power spectral density of five
individual one-dimensional trajectories 
and the ensemble average obtained from $150$
trajectories as a function of $\ln f T$, compared to the theoretical
relation $S \sim 1/f^2$, Eq. \eqref{concept}. The inset shows an image
of an individual polystyrene bead. 
Lower panel: Distribution of the dimensionless amplitudes $A^{(k)}$ 
for $k=1$ (black circles) and $k=2$ (blue triangles), obtained  from 150 bead trajectories 
in the range $50 \le f T \le 400$. The red solid curves depict the theoretical predictions $P(A^{(k)}=A)$
in Eq.~\eqref{limiting} for $k=1$ and $k=2$. The error bars are less than the size of the symbols.
\label{BEADS}}
\end{figure}

Lastly, we plot in Fig. \ref{BEADS} (upper panel) a logarithm of a
single-trajectory periodogram versus the logarithm of $f$, for five
trajectories of polystyrene beads in experiments performed within a flow cell
(see below for more details). Each trajectory represents a one-dimensional 
projection of the bead motion, which is a Brownian motion with the diffusion coefficient
$D \approx 0.365 \, {\rm um^2/s}$, as estimated from the bead's squared displacement averaged over $150$ trajectories. Note that the value of $D$ inferred from the averaged PSD appears to be pretty close, $D \approx 0.373 \, {\rm um^2/s}$. 
Again, we observe that the periodograms of the
bead motion agree very well with our prediction in Eq. \eqref{concept}, and
clearly follow the $1/f^2$-dependence for each individual trajectory. 
We stress again the similarly to the behavior of the MSD \eqref{tamsd}
\cite{pccp,thiel} to the amplitude scatter of a single-trajectory PSD as
function of $f$: apart from some small fluctuations, it is remarkably constant.

In the lower panel in Fig. \ref{BEADS}, using a set of $150$ individual
experimental trajectories,  we construct the distribution of the
amplitude $A$ in the $1/f^2$-dependence and compare it against our prediction
in Eq. \eqref{limiting}. The distributions of both one- and two-dimensional 
motions are shown, i.e., $k=1$ (projection along a line) and $k=2$ (projection along the imaging plane), 
respectively. Specifically, the distributions of the amplitudes are constructed by computing $A=f^2 S(f)/(4 D)$ for 350 
frequencies in the range $50 \le f T \le 400$. We note that the agreement with Eq. 
\eqref{limiting} is quite impressive for both $k=1$ and $k=2$, with the size of the error bars being less than the size of the symbols.

In our experimental setup, we used a flow cell
made of a cover slip and a plastic slide. Two holes were drilled in the plastic
slide to form the inlet and outlet of the chamber. Then the plastic slide
and cover slip were attached using double-sided tape and sealed with nail
polish, which was left to cure for $24$ hours. $1.2$-$\mu$m polystyrene beads
(SVP-10-5, Spherotech, Lake Forest, IL) were diluted in phosphate-buffered
saline (PBS) with $0.05$\% Tween $20$. The sample was agitated, centrifuged
and resuspended in PBS with $1$\% bovine serum albumin (BSA) and $0.05$\%
Tween $20$. Subsequently the bead suspension was introduced in the flow cell
chamber and the holes were sealed with nail polish, followed by immediate
imaging. The beads were imaged in an inverted microscope equipped with a
40$\times$ objective (Olympus PlanApo, N.A. $0.95$) and a sCMOS camera (Andor Zyla
4.2) operated at 100 frames per second. Each recorded sequence consisted of
$4,096$ frames. Bead tracking in the plane was performed in LabView using a
cross-correlation based tracking algorithm \cite{gosse2002magnetic}. An image
of a polystyrene bead is shown in the inset of Fig. \ref{BEADS}. Several
rings are observed in the image because the bead is intentionally not in
focus, in order to increase tracking accuracy.

We close this subsection with the following remark. Estimating the
diffusion coefficient or the temporal evolution of the mean-squared
displacement, one often slices a long trajectory into many smaller ones,
which permits to create some statistical average. In this case, it seems
to be appropriate for a stationary process because one here estimates either
a constant (diffusion coefficient) or deals with a monotonically growing
function of time. For the PSD the situation is much more subtle, because in
the discrete-time case it is a periodic function of the parameter $\omega$, with
an oscillatory behavior within each prime period. Therefore, in this case one
needs to fit the PSD only on a particular interval of variation of $\omega$
described below our Eq. \eqref{freq}. While slicing the trajectories permits one
to create some statistical ensemble here, as well, this occurs at the expense of
shrinking the interval $[\omega_l,\omega_r]$, and it is not clear \textit{a
priori\/} if such a procedure can be beneficial or detrimental to the analysis.
This question will be studied in detail elsewhere.

\subsection{Discrete-time case: Moment-generating function and distribution
of a single-trajectory periodogram}
\label{ZZ}

The expression  in Eq. \eqref{disc} permits us to 
write formally the moment-generating function and the distribution of a single-trajectory 
periodogram as
\begin{widetext}
\begin{subequations}
\begin{align}
\label{PhiR}
\Phi_{\lambda}\left(R_M(f) = R\right) &\equiv \mathbb{E}\left\{ \exp\left( - \lambda R_M(f)\right)\right\} = \frac{1}{2^M} \sum_{\{s_j\}}
\exp\left( - \frac{2 D \Delta^2 \lambda}{M} \left[\left(\sum_{j=1}^M  a_j s_j\right)^2 + \left(\sum_{j=1}^M b_j s_j\right)^2\right]\right)
\end{align}
and
\begin{align}
\label{distR}
P\left(R_M(f) = R\right) = \frac{1}{2^M} \sum_{\{s_j\}} \delta\left(R - \frac{2 D \Delta^2}{M} \left[\left(\sum_{j=1}^M a_j s_j\right)^2 + \left(\sum_{j=1}^M b_j s_j\right)^2\right]\right),
\end{align}
\end{subequations}
\end{widetext}
where the summation extends over all possible realizations of the sequence
of the increments $s_j$. These expressions are exact, but they are of a
little use, except for the (uninteresting) case when $M$ is
sufficiently small such that they can be written down in an explicit form
by enumerations of all possible sequences of $s_j$.

We notice next that the weighted sums 
\begin{subequations}
\begin{align}
\label{WM}
W_M = \sum_{j=1}^M  a_j s_j = \sum_{j=1}^M \left(\sum_{m=j}^M \cos\left(f \Delta m\right)\right) s_j \,,  
\end{align}
and
\begin{align}
\label{VM}
V_M = \sum_{j=1}^M  b_j s_j = \sum_{j=1}^M \left(\sum_{m=j}^M \sin\left(f \Delta m\right)\right) s_j \,,
\end{align}
\end{subequations}
appearing in Eqs. \eqref{PhiR} and \eqref{distR} can be considered as  two
coupled discrete-time random walks with variable, generally irrational and
incommensurate step-length.  From this observation, some general conclusions
can be reached about the expressions in Eqs. \eqref{PhiR} and \eqref{distR}:
(i.) For arbitrarily large $M$, both $W_M$ and $V_M$ are bounded from above
by their maximal displacements, meaning that $P\left(R_M(f) = R\right)$
has a bounded support and there is an upper cut-off value $R_{\rm max}(M)$
above which $P\left(R_M(f)\right)$ is identically zero. At the same time
(apart from some special cases, for instance, when  $f \Delta/\pi$ is a
rational number, see Appendices \ref{A} and \ref{B}) the trajectories $W_M$
and $V_M$, for whatever large $M$, will never visit the origin again which
signifies that $P\left(R_M(f) = 0\right) \equiv 0$. (ii.) Moreover, we may
expect that for an arbitrarily large $M$ there is some gap $R_{\rm min}(M)$
below which $P\left(R_M(f)\right)$ is identically zero too, that is, $W_M$
and $V_M$ do not visit simultaneously some vicinity of the origin. These two
specific features of the discrete-time settings make a significant difference
as compared to the truncated Wiener's representation, in which analogous sums
$\sum_{n=1}^N g_n \zeta_n$ and $\sum_{n=1}^N h_n \zeta_n$ are not bounded,
and may, in principle, be equal to zero, since the increments $\zeta_n$
are continuous random variables with the support $(-\infty,\infty)$.

We proceed with the derivation of closed-form expressions for the
moment-generating function and for the distribution of a single-trajectory
periodogram.  Using the integral identity in Eq. \eqref{identity}, we cast
Eqs.\eqref{PhiR} and  \eqref{distR} into the form which permits to perform
the averaging very directly. This leads to
\begin{widetext}
\begin{subequations}
\begin{align}
\label{dis_phi}
\Phi_{\lambda}\left(R_M(f)\right) &= 
\frac{M}{8 \pi D \Delta^2 \lambda}  \int^{\infty}_{-\infty} dx  \int^{\infty}_{-\infty} dy \exp\left(- \frac{M}{8 D \Delta^2 \lambda} \left(x^2 + y^2\right)\right) \prod_{j=1}^M \cos\left(x a_j + y b_j\right)
\end{align}
and
\begin{align}
\label{dis_p}
P\left(R_M(f) = R\right) &= \frac{M}{8 \pi D \Delta^2} \int^{\infty}_{-\infty}   \int^{\infty}_{-\infty} dx \, dy  \, 
J_{0}\left(\sqrt{\frac{M \left(x^2 + y^2\right) R}{2 D \Delta^2}}\right) \prod_{j=1}^M \cos\left(x a_j + y b_j\right),
\end{align}
\end{subequations}
\end{widetext}
where $J_0(\ldots)$ is the Bessel function of the $1$st  kind.
Expressions  \eqref{dis_phi} and \eqref{dis_p} are formally exact and
hold for arbitrary $M$, $f$ and $\Delta$.  We note that there is a set
of \emph{magic frequencies\/} $f$ such that $f \Delta = \pi q$ with $q$
a rational number, when $a_j$ and $b_j$ become periodic functions of $j$
and the kernel  $\prod_{j=1}^M \cos\left(x a_j + y b_j\right)$ can be written
down explicitly in form of a finite series of cosines. Then, the integrations
in Eqs. \eqref{dis_phi} and \eqref{dis_p} can be performed exactly resulting
in exact closed-form expressions for $\Phi_{\lambda}\left(R_M(f)\right)$
and $P\left(R_M(f) = R\right)$.  Several examples of such calculations
are presented in Appendices \ref{A} and \ref{B}. However, for arbitrary
$f$, analytical calculation of the integrals in results \eqref{dis_phi}
and \eqref{dis_p} is certainly beyond reach and one has to resort to some
approximations.

\subsection{Discrete-time case in the limit $M \gg 1$: Limiting forms of the moment-generating function and of the distribution of a single-trajectory periodogram}

\begin{figure}
\includegraphics[width=8.8cm]{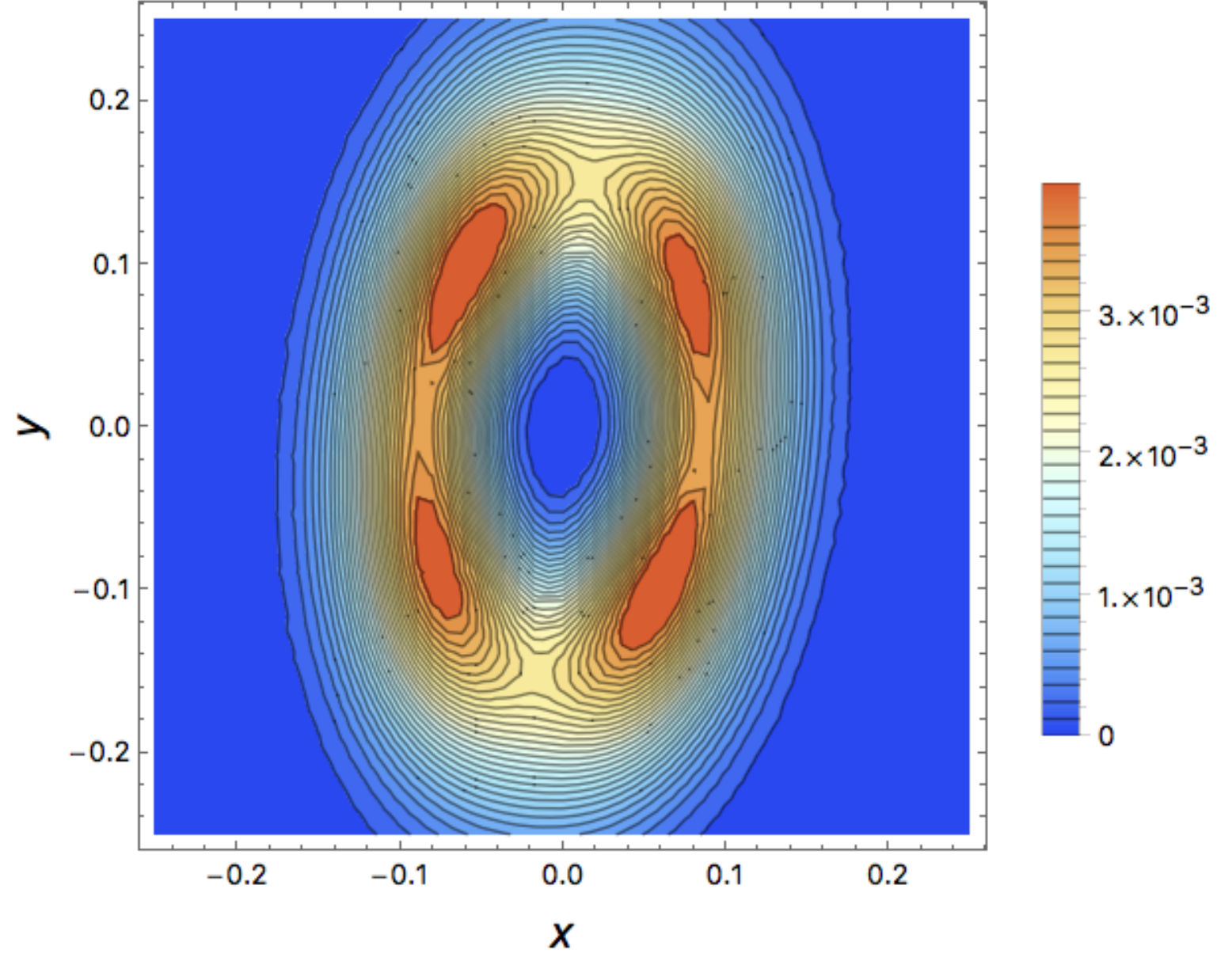}
\includegraphics[width=8.8cm]{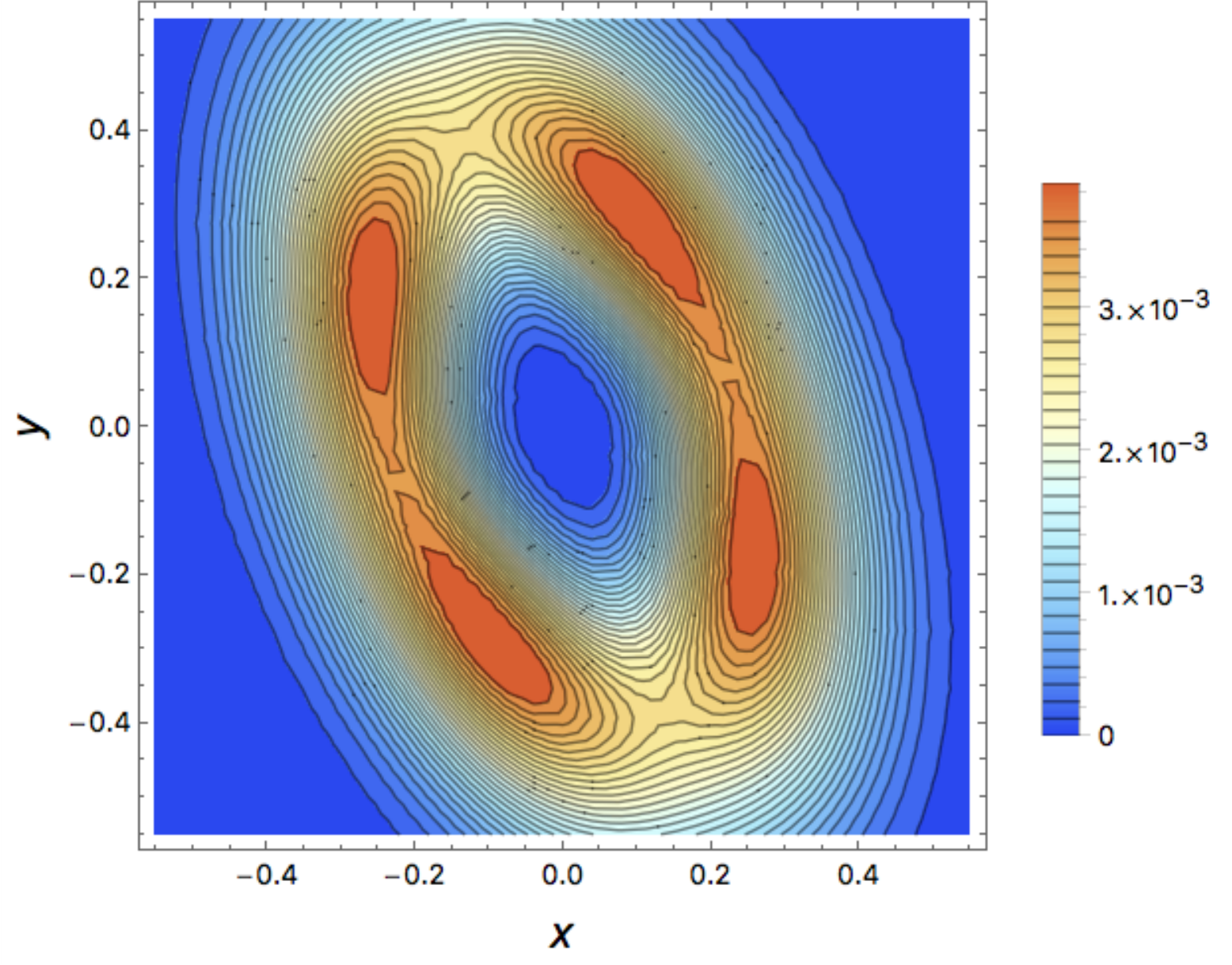}
\caption{Density plot of $\kappa(x,y)$ (see the text below Eq. \eqref{app}).
Upper panel:  $M = 10^2 $ and $f T = 12$.
Lower panel:   $M = 10^2 $ and $f T = 120$.
\label{FIG15}
}
\end{figure} 

Meaningful approximations of expressions \eqref{dis_phi} and \eqref{dis_p} can be obtained 
 in the large-$M$ limit. One has, however, to distinguish between the case
 when $f T$ is fixed and $M \to \infty$, and the case when $M \to \infty$
but $f T$ scales with $M$ and may take any value within the interval $(0, 2 \pi M)$.

\subsubsection{Fixed $f T$ and $M \to \infty$}
 
 For fixed $T$ and not too large $\lambda$, the parameter $M/(8 D \Delta^2 \lambda) = M^3/(8 D T^2 \lambda)$ is large 
 when $M \gg 1$, so that the exponential function in Eq. \eqref{dis_phi} is small everywhere except for the vicinity of the origin. 
 Moreover,  the integrals 
 in Eqs. \eqref{dis_phi} and \eqref{dis_p}
look very similar to the ones appearing in the theory of Pearson's random 
walks with a variable step length 
(see, e.g., Ref. \cite{hughes}) and involve a kernel $\prod_{j=1}^M \cos\left(x a_j + y b_j\right)$.
It is well known that in the limit $M \to \infty$ the behavior of such a  kernel  
is determined overwhelmingly by its behavior in the vicinity of $x =0$ and $y=0$, so that one has
\begin{align}
\label{app}
\prod_{j=1}^M \cos\left(x a_j + y b_j\right) \sim \exp\left(- \frac{1}{2} \sum_{j=1}^M \left(x a_j + y b_j\right)^2\right) \,.
\end{align}
The accuracy of this approximation is checked in Fig. \ref{FIG15} in which we depict a density plot of the difference $\kappa(x,y)$
of  
$\prod_{j=1}^M \cos\left(x a_j + y b_j\right)$ and of the exponential function in the right-hand-side of \eqref{app}. 
Already for $M = 10^2$ the difference 
$\kappa(x,y)$ is zero almost everywhere and in the regions where it deviates from zero it is nonetheless numerically very small. 

Next, we have that for $f T$ fixed and $M \gg1$ the ensemble-averaged periodogram
and its variance admit the following representations
\begin{subequations}
\begin{align}
\label{z30}
\mu_R(f,T) = \mu_S(f,T) + \frac{4 D \sin^2\left(f T/2\right)}{f^2} \frac{1}{M} + O\left(\frac{1}{M^2}\right)
\end{align}
and
\begin{align}
\label{z31}
&\sigma^2_R(f,T) = \sigma^2_S(f,T) + \frac{D^2}{f^4}\Bigg(4 \left(\cos\left(2 f T\right) - 6 \cos\left(f T\right) - 7\right) \nonumber\\
&+ 8 \left(7 - \cos\left(f T\right) \frac{\sin\left(f T\right)}{f T}\right)\Bigg) \frac{1}{M} + O\left(\frac{1}{M^2}\right),
\end{align}
\end{subequations}
where the omitted terms have a bounded amplitude for any value of $f T$ and
decay with the growth of $M$ in proportion to the second inverse power of $M$.  

Inserting expression \eqref{app} into Eqs. \eqref{dis_phi} and \eqref{dis_p} and performing the integrations, we 
find that in the limit $M \gg 1$,
for fixed $f T$ and bounded $\lambda$, 
the moment-generating function 
and the distribution of a single trajectory periodogram
are given 
up to terms of order $O\left(1/M\right)$,
by our previous results \eqref{MG} and \eqref{dist}.

\subsubsection{Arbitrary $f T \in (0, 2 \pi M)$ and $M \to \infty$.}

We now 
relax the condition that $f T$ is fixed, and suppose that it can attain any value within the prime period $2 \pi M$. In other words, we consider the situation in which $f T$ scales with $M$ such that the expansions  \eqref{z30} and \eqref{z31} are invalid, and we can no longer discard the correction terms. In this case, 
 we find that the moment-generating function of a single-trajectory periodogram is given by
\begin{align}
\label{67}
&\Phi_{\lambda}\left(R_M(f)\right) = \Bigg[1 + \frac{4 D \Delta^2 \lambda}{M} \left(\sum_{j = 1}^M \left(a_j^2 + b_j^2\right)\right) \lambda \nonumber\\
&+ \frac{16 D^2 \Delta^4}{M^2} \left( \left(\sum_{j, i = 1}^M a_j^2  b_i^2\right)
- \left(\sum_{j=1}^M a_j b_j\right)^2\right) \, \lambda^2\Bigg]^{-1/2} \,.
\end{align}
Differentiating this expression once and twice, and then setting $\lambda = 0$, we observe that \eqref{67} correctly reproduces the first moment of the periodogram, but yields an incorrect expression for the variance, meaning that in this limit the approximation \eqref{app} is insufficient (although it works fairly well in the previous case with $f T$ kept fixed). In other words, 
in order to obtain a moment-generating function which reproduces correctly first two moments of a single-trajectory periodogram, 
one has to go beyond 
this approximation. 
For Pearson's random walks with  $j$-dependent step-lengths $a_j$ and $b_j$ 
this turns out to be quite a complicated 
problem which we are not in the position to solve here. 

Conversely, on intuitive grounds, we may conjecture such a form of the
moment-generating function which reproduces the first and the second moment
correctly. For $M \to \infty$, this is given by
\begin{align}
\label{period}
\Phi_{\lambda} &= \Bigg[1 + 2 \mu_R(f, T) \lambda + 
 \left(2  - \gamma^2_R\right) \mu_R^2(f, T) \, \lambda^2\Bigg]^{-1/2} \,,
\end{align}
that is, has precisely the same form as the moment-generating function in
the continuous-time case, Eq. \eqref{MG}, but with the first two moments
 replaced by the analogous properties of a single-trajectory
periodogram. Evidently, a generalization to the case of a $k$-component
periodogram amounts to merely replacing $1/2$ by $k/2$.

In turn, the form \eqref{period} 
implies that the distribution of a single-trajectory periodogram is given by Eq. \eqref{dist} with $\mu_S(f,T)$ replaced by $\mu_R(f,T)$ and $\sigma_S^2(f,T)$ replaced by its discrete-time counterpart $\sigma^2_R(f,T)$, that is,
\begin{widetext}
\begin{align}
\label{dist10}
P\left(R_M(f)=R\right)&=\frac{1}{\sqrt{2-\gamma^2_R}\mu_R(f,T)}\exp\left(-\frac{
1}{2-\gamma^2_R}\frac{R}{\mu_R(f,T)}\right){\rm I}_0\left(\frac{\sqrt{\gamma^2_R
-1}}{2-\gamma^2_R}\frac{R}{\mu_R(f,T)}\right).
\end{align}  
\end{widetext}
We are unable to prove expression \eqref{period} (and hence,
Eq.~\eqref{dist10}) in the general case of arbitrary $f$.  However, as we have
already remarked, there are plenty of cases in which the moment-generation
function and the distribution in Eqs. \eqref{dis_phi} and
\eqref{dis_p} can be calculated exactly.  In Appendix \ref{B} we present
such calculations for several particular values of the frequency, such that
$f \Delta = 2 \pi/3$, $f \Delta = \pi/2$, $f \Delta = 2 \pi$ and $f \Delta =
\pi$, and show that the exact discrete-time results do indeed converge to
the asymptotic forms \eqref{period} and \eqref{dist10}. Therefore, we have
all grounds to believe that the expression \eqref{dist10} is correct.

Lastly, we note that Eq. \eqref{dist10} permits us to make a substantial generalization of the single-trajectory relation \eqref{concept}. We see that for $f T \in (\omega_l, \pi M - \omega_l)$ (in which domain $\gamma_R = \sqrt{5}/2$ with any necessary accuracy set by the choice of $\varepsilon$), a single-trajectory periodogram obeys
\begin{align}
\label{concept2}
R_M(f) = A \, \mu_R(f,T) \,,
\end{align}
implying that the spectrum of a single-trajectory periodogram should be the same as of the ensemble-averaged periodogram,  
while only the amplitude $A$ will be a fluctuating property. The limiting 
distribution of the amplitude is given by Eq. \eqref{limiting}.

\section{Conclusions and Discussion}
\label{conc}

We studied here in detail the statistical properties of the power spectral
density of a single trajectory of a $d$-dimensional Brownian motion. We
calculated exactly the moment-generating function, the full probability
density function and the moments of arbitrary, not necessarily integer order
of such a PSD in the most general case of arbitrary frequency $f$, arbitrary
(not necessarily infinite) observation time $T$ and arbitrary number of
projections of a given trajectory onto the coordinate axes. We showed that
for a sufficiently large $T$ (and the frequency $f> 0$) a single-trajectory
PSD for any realization of the process is proportional to its first moment,
which embodies the full dependence on the frequency and on the diffusion
coefficient.  This implies that the correct frequency-dependence specific to
an ensemble of trajectories can be deduced already from a single trajectory
and solely the numerical proportionality factor in this relation is random,
and varies from realization to realization.  Due to this fact, one cannot
infer the precise value of the ensemble-averaged diffusion coefficient from a
single-trajectory PSD since the ensemble-averaged PSD is linearly proportional
to the diffusion coefficient which thus appears to be multiplied by a random
amplitude.  The distribution function of this amplitude was also calculated
exactly here and its effective width was quantified using standard criteria.

Moreover, we addressed several questions emerging in connection with the
numerical and experimental verification of our analytical predictions for the
continuous-time BM. To this end, we first considered Wiener's  representation
of BM in form of an infinite Fourier series with random coefficients, whose
truncated version is often used  in numerical simulations. We showed that
the distribution of the single-trajectory PSD  obtained from Wiener's series
in which just $N$ terms are kept, instead of an infinite number, has exactly
the same form as the one obtained for the continuous-time BM when $f T$ is
within the interval $(0, \pi N)$.  Outside this interval, the probability
density function of the truncated PSD converges to a different form. Next,
we examined the case when a trajectory of continuous-time BM is recorded at
some discrete time moments, so that the whole trajectory is represented by
a set of $M$ points and the PSD, called a periodogram, becomes a periodic
function of the product $f T$ with the prime period equal to $2 \pi M$.
We analyzed several aspects of this discrete-time problem. Namely, we studied
how big $M$ should be taken at a fixed observation time $T$ so that we may
recover the results obtained for the continuous-time BM. Apart of that,
we studied the limiting forms of the distribution of a single-trajectory
periodogram and showed, in particular, that for $f T$ kept fixed and $M \to
\infty$, the latter converges to the form obtained for the continuous-time
BM. In contrast, when $f T$ is left arbitrary so that it may assume any
value within the prime period, that is, $f T \in (0, 2 \pi M)$, the limiting
distribution of  a single-trajectory periodogram converges to a different
form as $M \to \infty$. Our analysis revealed the remarkable observation
that for $f T$ belonging to a certain interval within the prime period,
a single-trajectory periodogram equals, up to a random numerical amplitude,
the ensemble-averaged periodogram, and the latter embodies the full dependence
on $f$ and $T$. Therefore, similarly to the continuous-time case, the correct
spectrum can be obtained already from a single trajectory.

To check our analytical predictions we performed numerical analysis, using
a truncated Wiener's representation and also Monte Carlo simulations of
discrete-time random walks, as well as experimental analysis of Brownian
motion of polystyrene beads in aqueous solution. We confirmed our prediction
of the relation connecting  a single-trajectory PSD and its ensemble averaged
counter-part and demonstrated that the former indeed exhibits a well-defined
$1/f^2$-dependence, specific to a standard, ensemble-averaged PSD. Furthermore,
our numerical analysis confirmed the form of the distribution function of
the amplitude in this relation.

The theoretical analysis for the case of Brownian motion presented here
can be extended in several directions. In particular, one may inquire about
analogous distributions for anomalous diffusion processes \cite{pccp}, as
exemplified, for instance, by superdiffusive L\'evy motion or subdiffusive
continuous-time random walks with a broad distribution of the waiting times.
Another important example of transport processes is given by a wide and
experimentally relevant class of anomalous diffusion processes called
fractional Brownian motion, for which the generalization of our analysis is
relatively straightforward. Subdiffusive fractional Brownian motion is related
to the overdamped, generalized Langevin equation with a power-law memory kernel
typical for viscoelastic systems \cite{goychuk}. However, superdiffusive
fractional Brownian motion was identified for active transport in biological
cells \cite{reverey}. We also mention the very active field of diffusion
with varying diffusion coefficients. These can be systematically varying in
space \cite{hdp}, time \cite{lim,anna}, or be randomly varying in time
\cite{gary,klsebastian,chechkin} or space \cite{lapeyre,lapeyre1,andrey}. Lastly, a
challenging field of further research is to search for such periodic functions
of the frequency in Eq. \eqref{spec0}, in place of the exponential function, for
which a single-trajectory PSD will possess an ergodic property so that its
variance will tend to zero in the limit $T\to\infty$.

\section*{Acknowledgments}

DK, RM and GO wish to thank for warm hospitality the International Center
for Advanced Studies in Buenos Aires, where this work was initiated. DK
acknowledges a partial support from the NSF Grant No. 1401432. EM acknowledges
funding from the European Research Council (ERC) under the European Union
Horizon 2020 research and innovation program (grant agreement No 694925).

\appendix

\section{\label{A} Moment-generating function and the distribution of the
PSD for $f=0$}

We focus in this Appendix on the moment-generating function and on
the probability density function of a single-trajectory PSD for the
continuous-time BM and of the discrete-time single-trajectory periodogram
in the special case $f=0$.

Consider first the case of BM. 
The first moment and the variance of the PSD for $f =0$ follow from Eqs. \eqref{power_av}  and \eqref{var} 
and are given explicitly by
\begin{align}
\label{atperiodcont}
\mu_S(0,T) = \frac{2 D T^2}{3} \,,
\end{align} 
and
\begin{align}
\label{atperiodcont1}
\sigma_S^2(0,T) = \frac{8 D^2 T^4}{9}  \,,
\end{align}
so that 
 the coefficient of variation is $\gamma_S \equiv \sqrt{2}$.  Hence, in this case
 the coefficient in front of the quadratic term in Eq. \eqref{MG} vanishes and the moment-generating function obeys
\begin{align}
\label{MGf0}
\Phi_{\lambda}\left(\tilde{S}^{(k)}_T(0)\right) &= \Bigg[1 + \dfrac{4 D T^2}{3} \, \lambda \Bigg]^{-k/2} \,.
\end{align} 
Taking the inverse Laplace transform, we then find  that in this case
the distribution  $P(\tilde{S}_T^{(k)}(0))$ is given by 
\begin{align}
\label{distf0}
P(\tilde{S}_T^{(k)}(0)=S) &= \left(\frac{3}{4 D T^2}\right)^{k/2} \frac{S^{k/2 - 1}}{\Gamma(k/2)}  \exp\left(- \frac{3 S}{4 D T^2}\right)
 \,,
\end{align} 
that is, it is the $\chi^2$-distribution with $k$ degrees of freedom. Note that in contrast to the distribution \eqref{dist} with $f$ bounded away from zero, 
the one in Eq. \eqref{distf0} does not attain 
a limiting form as the observation time $T \to \infty$. For any fixed $S > 0$ one has
\begin{align}
P(\tilde{S}_T^{(k)}(0)=S) \sim \frac{1}{T^k} \,, \,\,\, \text{as $T \to \infty$} \,.
\end{align}

\begin{figure}
\includegraphics[width=8.5cm]{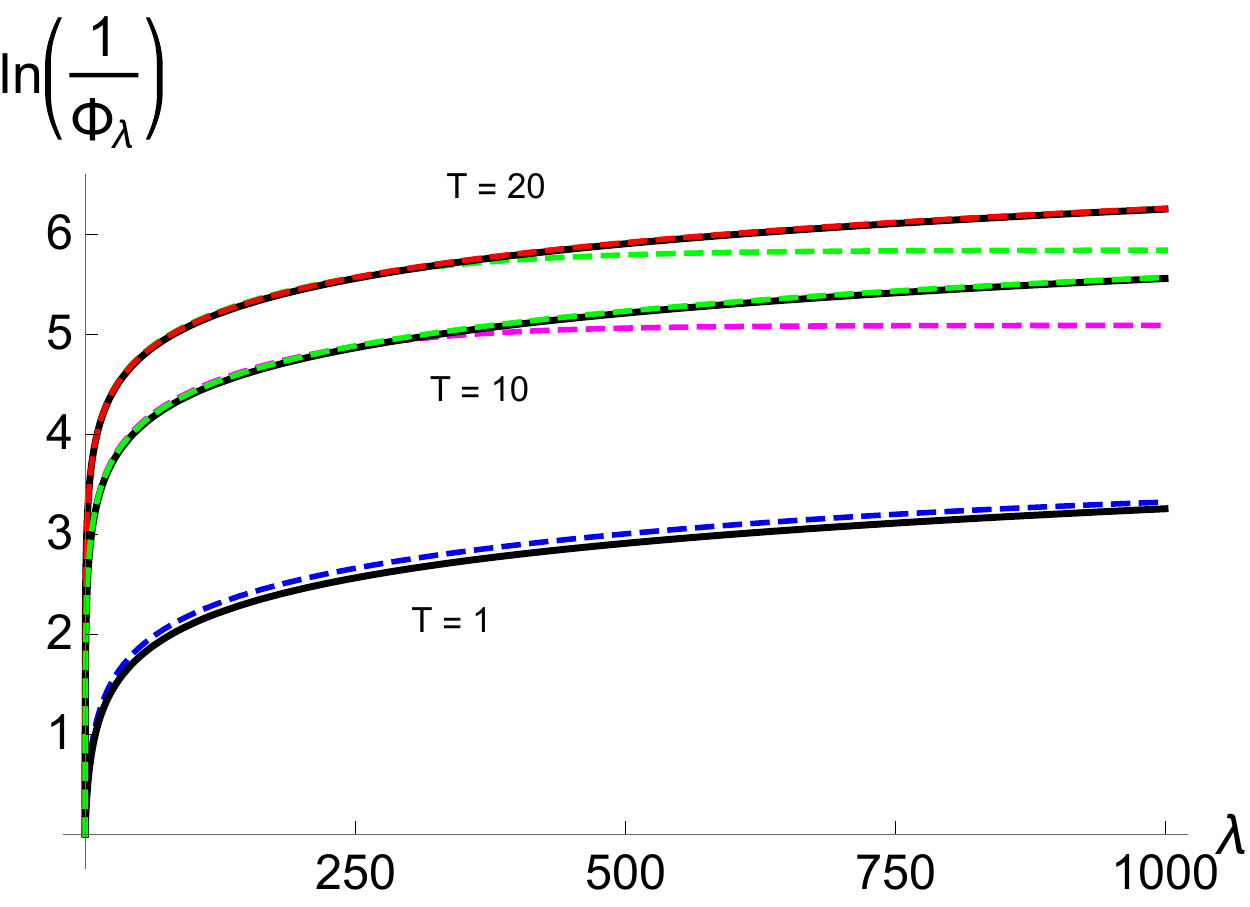}
\includegraphics[width=8.5cm]{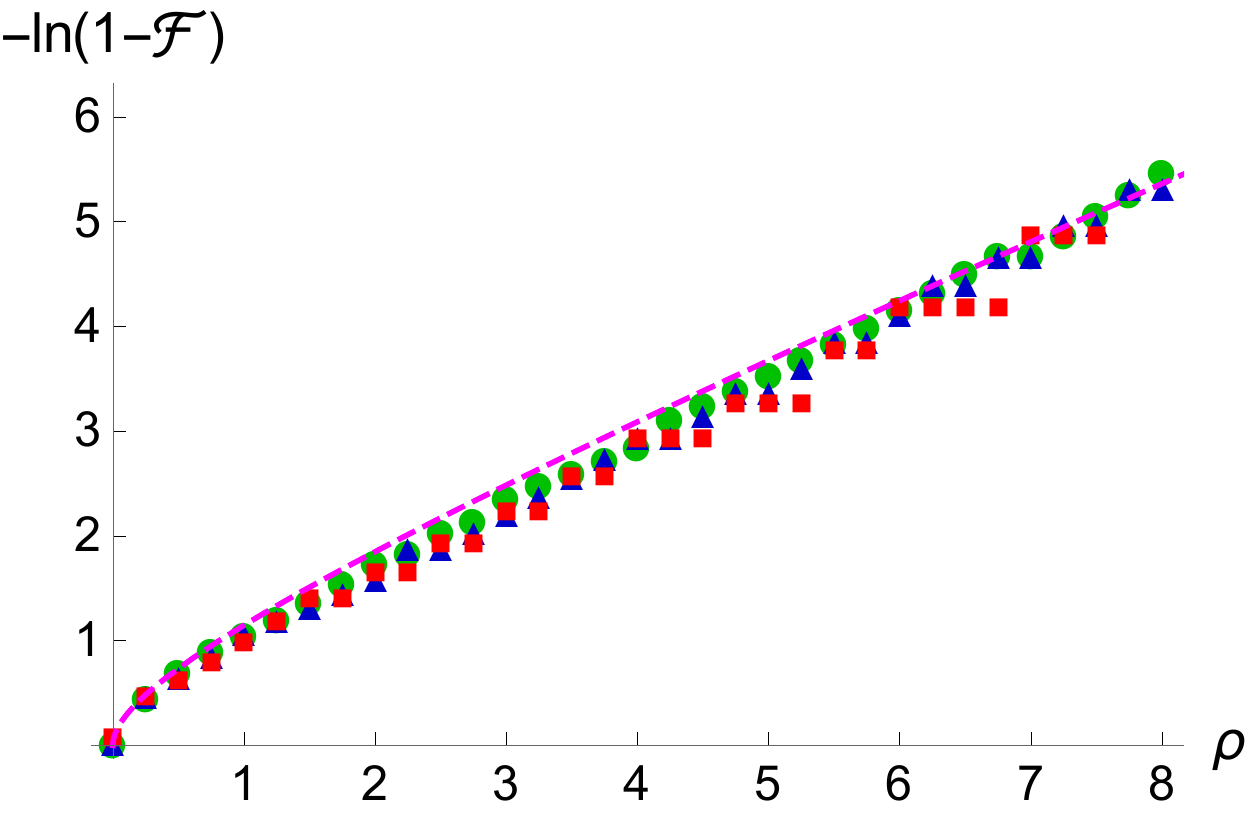}
\caption{The case $f = 0$.
Upper panel: Logarithm of $\Phi_{\lambda}$ versus  $\lambda$ for the continuous-time case \eqref{MGf0} with $k=1$, and for the discrete-time case \eqref{MGdisf0}. 
The diffusion coefficient is $D = 1/2$. The solid curves from top to bottom correspond to the result \eqref{MGf0} for $T = 20$, $T = 10$ and $T = 1$, respectively. The dashed curves are the results for the discrete-time case with different $T$ and $M$: $T = 1$ and $M = 12$ (blue), $T= 10$ and $M = 36$ (magenta),  $T= 10$ and $T = 20$ with $M = 60$ (green) and $ T= 20$ with $M = 100$ (red).  Lower panel: 
Logarithm of $1 - {\cal F}$, ${\cal F}$ being the cumulative distribution function, versus variable $\rho = R/\mu_R(0,T)$ for $T = 1$.
Dashed curve (magenta) gives the cumulative distribution function for the continuous-time $\chi^2$-distribution \eqref{distf0}. Symbols  - squares ($M = 8$),  triangles ($M = 12$) and  circles ($M = 16$) - 
correspond to ${\cal F}$ for the discrete-time distribution \eqref{DISTdiscf0}.   
\label{FIG10}
}
\end{figure} 

In the discrete-time case, for $f=0$, 
the first moment $\mu_R(0,T)$ of the periodogram, the variance $\sigma^2_R(0,T)$ and the variation coefficient $\gamma_R$
are given by Eqs. \eqref{atperiod}, \eqref{atperiod1} and \eqref{atperiod2}, respectively. We notice that they converge, as $M \to \infty$, to their continuous-time counterparts in Eqs. \eqref{atperiodcont} and \eqref{atperiodcont1}, and $\gamma_R \to \gamma_S \equiv \sqrt{2}$, meaning that 
in this limit the moment-generating function and the distribution of the periodogram converge  to the forms in Eqs. \eqref{MGf0} and \eqref{distf0}.

For $f =0$ in the discrete-time case, the moment-generating function and the distribution of the periodogram can be calculated exactly for arbitrary $M$, 
which permits us to estimate the rate of convergence.
 We constrain our analysis to the case 
when $M$ is divisible by $4$.
We impose such a constraint just for a simplicity of exposition, exact solutions can be also obtained 
for other values of $M$
resulting, however, in rather cumbersome expressions without providing any additional insight. 

We notice that for $f=0$ 
the coefficients $b_j$ in Eq. \eqref{bj} are identically equal to zero, which implies that $V_M \equiv 0$, \eqref{VM},
while the coefficients $a_j$ in Eq. \eqref{aj} are explicitly given by $a_j \equiv M + 1 - j$. Consequently, the sum $W_M = \sum_{j=1} (M + 1 - j) s_j$ appearing in Eqs. \eqref{PhiR} and \eqref{distR} can be thought of as a "random walk" with a linearly "shrinking" time-dependent step length (see, e.g., Refs. \cite{red1,red2} for other examples of such random walks). In this case, the product of cosines in Eqs. \eqref{dis_phi} and \eqref{dis_p} can be explicitly written down as 
\begin{align}
\label{exp7}
&\prod_{j=1}^M \cos\left(a_j x + b_j y\right) = \prod_{j=1}^M \cos\left(\left(M + 1 - j\right) x\right) = \nonumber\\
&=\frac{1}{2^M} 
 \sum_{m= - M(M+1)/4}^{M(M+1)/4} q_M\left(\frac{M(M+1)}{4} - m\right) \cos\left(2 m x\right) \,,
\end{align}
where the coefficient $q_M(j)$ is defined by
\begin{align}
q_M(j) = \frac{1}{j!} \left. \frac{d^j}{dq^j} \left(-q,q\right)_M\right|_{q = 0} \,,
\end{align}
with $(-q,q)_M = \prod_{m=1}^M (1+q^m)$ being the $q$-Pochhammer symbol \cite{ernest}. In other terms, $q_M(j)$ is the 
numerical 
coefficient before the term $q^j$ in the polynomial $\prod_{m=1}^M (1+q^m)$. 
For $M=\infty$, $q_{\infty}(j)$ is simply the number of all possible partitions of an 
integer $j$ into \textit{distinct} parts. Note also that  $q_M(j)$ is symmetric around $m=0$
such that $q_M\left(M(M+1)/4 - m\right) = q_M\left(M(M+1)/4 + m\right) $, $m = 1, 2, \ldots, M(M+1)/4$.

 Inserting next expression \eqref{exp7} into Eqs. \eqref{dis_phi} and \eqref{dis_p} and performing the integrations, we arrive at the following exact results for the moment-generating function and the distribution of a single-trajectory periodogram $R_M(f)$  for $f = 0$:
\begin{align}
\label{MGdisf0}
\Phi_{\lambda} &= \frac{1}{2^M}  \sum_{m= - M(M+1)/4}^{M(M+1)/4} q_M\left(\frac{M(M+1)}{4} - m\right) \nonumber\\
& \times \exp\left( - \frac{8 D \Delta^2 m^2}{M} \lambda\right) \,,
\end{align} 
and
\begin{align}
\label{DISTdiscf0}
&P(R_M(0) = R) = \frac{1}{2^M} \Bigg[  q_M\left(\frac{M(M+1)}{4}\right)  \delta\left(R\right) + \nonumber\\
&+ 2 \sum_{m=1}^{M(M+1)/4} q_M\left(\frac{M(M+1)}{4} - m\right) 
\delta\left(R  - \frac{8 D \Delta^2 m^2}{M} \right)
 \Bigg]\,.
\end{align}
Note that $R$ has a discrete support on the interval 
$(0,R_{\rm max})$ with $R_{\rm max} = D \Delta^2 M (M+1)^2/2$. Note, as well, that the presence of a delta-function at $R = 0$
is the consequence of the choice of $M$; for $M$ not divisible by $4$ the coefficient in front of $\delta(R)$ equals zero because
the "random walk" $W_M$ never visits the origin.

In Fig. \ref{FIG10}, upper panel,  
we present a comparison of the expressions for the moment-generating functions in Eq. \eqref{MGf0} for the continuous-time case
and in Eq. \eqref{MGdisf0} for the discrete-time case, for different densities of the discrete-time points for different values of $T$. We observe that for $T=1$ and $M=12$ (corresponding to $\Delta = 1/12$) the agreement between the continuous- and the discrete-time results is fairly good up to $\lambda = 10^3$.
For $T = 10$ and $M = 36$, such that on average
$3.6$ points are recorded for a unit of time, the discrete-time result agrees with the continuous-time one up to $\lambda = 3 \times 10^2$.  Increasing the number of points up to $M = 60$, so that we have now $6$ (instead of $3.6$) points per unit of time, a perfect agreement between Eqs. \eqref{MGf0} 
and \eqref{MGdisf0}  extends up to $\lambda = 10^3$. Lastly, for the observation time $T = 20$ with $60$ recorded points (which corresponds 
to just $3$ points per unit of time $T$) we have a perfect agreement between the discrete- and continuous-time predictions for values of $\lambda$ up to $\lambda = 3 \times 10^2$ and a significant departure for $\lambda$ exceeding these values. In contrast, for $T = 20$ and $M = 10^2$, such that 
we have $5$ points for each unit of time, a perfect agreement is observed for the whole range of variation of $\lambda$. We thus conclude that, in fact, 
once we seek to achieve a good agreement between the continuous- and discrete-time expressions for the moment-generation function, 
we do not need a very high precision in approximating the  
continuous-time trajectory by a discrete-time one.

Next,  in the lower panel of Fig. \ref{FIG10} we compare the probability density functions obtained for the continuous-time and the discrete-time cases. 
 Since the distribution  in the discrete-time case is a finite sum of delta-functions (see Eq. \eqref{DISTdiscf0}) it is appropriate 
 to compare not the distributions themselves, but rather their integrated forms, the cumulative distribution functions ${\cal F}$, obtained by integrating the forms in Eqs. \eqref{distf0} and \eqref{DISTdiscf0} over $d R$ from $0$ to $R$. 
From Fig. \ref{FIG10} (lower panel) we conclude 
that here too just $16$ recorded points per unit of time appear to be enough 
to get a fairly good agreement between
the distributions  in the continuous- and the discrete-time cases.

\section{\label{B} Several exact results for the moment-generating function and the distribution of
a single-trajectory periodogram}

In this Appendix we present several stray examples for which the
moment-generating function in Eq. \eqref{PhiR} and the distribution of
a single-trajectory periodogram in Eq. \eqref{distR} can be calculated
exactly. Our aim here is to demonstrate that in all these particular cases
the exact moment-generating function and the distribution converge  to the
forms \eqref{period} and \eqref{dist10} as $M \to \infty$, which validates
our conjecture.  These particular choices of the frequency within the prime
period are: $f \Delta = \pi$, $f \Delta = \pi/2$, $f \Delta = 2 \pi/3$
and the endpoint of the prime period, $f \Delta =  2 \pi$.  We note that
for all choices of the frequency the first moment $\mu_R(f,T)$ and the
variance $\sigma_R(f,T)$ converge as $M \to \infty$ to different values,
as compared to their continuous-time counterparts. This means, in turn, that
for $f T$ within the prime period these are the forms in Eqs. \eqref{period}
and \eqref{dist10} which describe the moment-generating function and the
distribution of the periodogram in the limit $M \to \infty$, but not the
continuous-time results derived in Sec. \ref{BM}.

\subsection{The case $f \Delta =  \pi$}

We start with the simplest case, where the convergence of  the
moment-generating function \eqref{PhiR} and of the distribution of a
single-trajectory periodogram \eqref{distR} to the limiting forms in
Eqs. \eqref{period} and \eqref{dist10} can be established analytically. We
will again assume, for simplicity, that $M$ is divisible by $4$. We hasten
to remark that such a constraint is not crucial---an exact solution can be
obtained in the general case of an arbitrary $M$, but the resulting expressions will be more cumbersome.

\begin{figure}
\includegraphics[width=8.5cm]{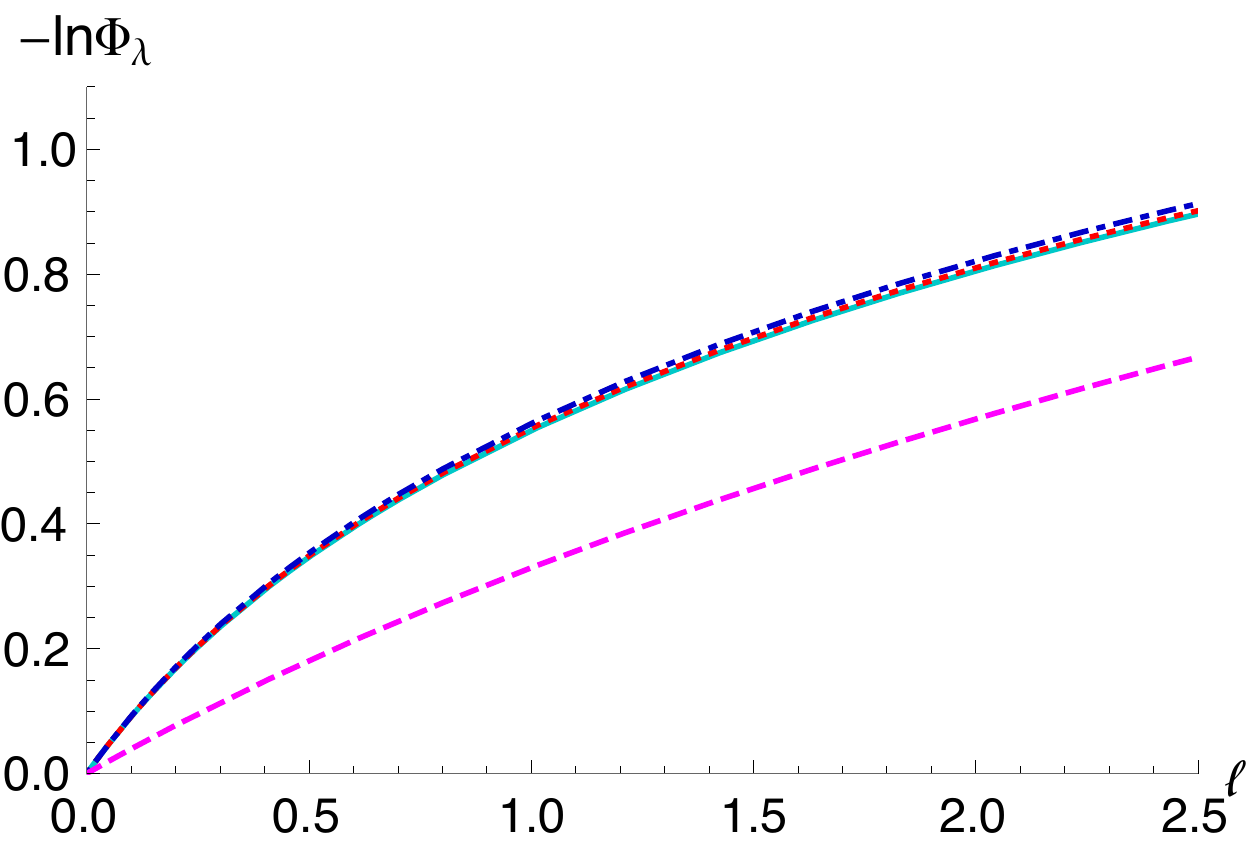}
\includegraphics[width=8.5cm]{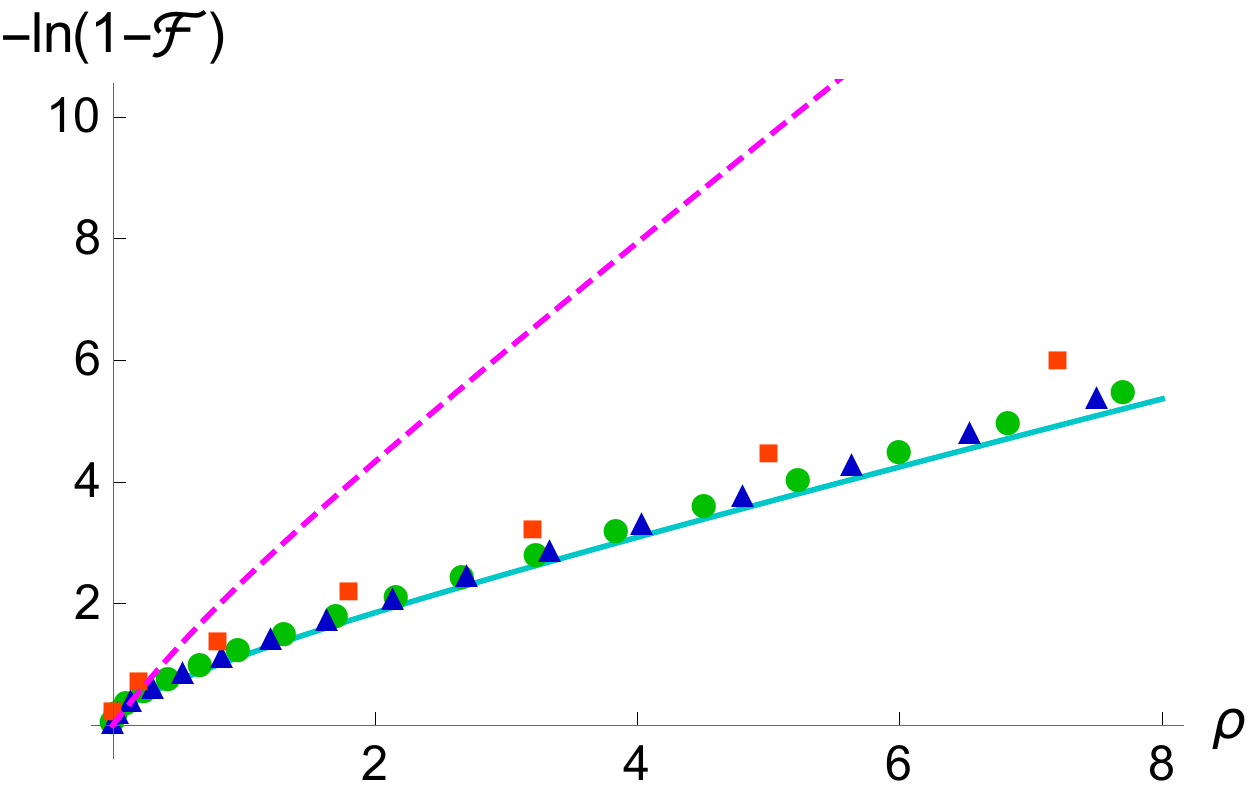}
\caption{The case $f \Delta = \pi$ (one-half of the prime
period).  Upper panel: The moment-generating function as a function of
$\textit{l} = \lambda/\mu_R$. The solid (cyan) curve shows the limiting form of the
moment-generating function in Eq. \eqref{period} (or equivalently, in
Eq. \eqref{1_10}). Dotted (red) and dot-dashed (blue) curves
represent the exact discrete-time moment-generating
function in Eq. \eqref{1_07} for $M = 60$ and  $M = 20$, respectively.
 The dashed curve depicts our continuous-time prediction
in Eq. \eqref{MG} with $f = \pi/\Delta$ and $T = 1$: it is presented here to demonstrate
that the moment-generating function obtained in the discrete-time case
does converge to the form  in Eq. \eqref{MG} but to a distinctly different 
form in Eq. \eqref{period}.  Lower panel: The cumulative distribution function ${\cal
F}$ versus $\rho = R/\mu_R$.  The solid curve represents the cumulative
distribution function of our conjectured distribution in Eq. \eqref{dist10}
(or equivalently, in Eq. \eqref{1_08}), while the symbols correspond to the
cumulative distribution function of the exact distribution in Eq. \eqref{1_08}:
$M = 40$ (squares), $M = 240$ (triangles)  and $M = 300$ (circles). The dashed
curve depicts the cumulative distribution function of the continuous-time
result in Eq. \eqref{dist} with $f = \pi/\Delta$ and $T = 1$.
\label{fig_plot1}}
\end{figure}

For $f \Delta = \pi$, the coefficients $b_j$ in Eq. \eqref{bj} all vanish, 
such that the "random walk" $V_M$ \eqref{VM} stays at the origin.  
The coefficients $a_j$ in Eq. \eqref{aj} are 
periodic with period $2$; that being, $a_{j} = a_{j+2}$, 
and are given by $a_j=(1 +(-1)^j)/2$ when $M$ is divisible by $4$.  Consequently, the random walk $W_M$ in Eq. \eqref{WM} 
steps at even time moments and pauses at the odd ones. 
In this case, the first moment and the variance of a single-trajectory periodogram are respectively given by 
Eqs. \eqref{peaks1} and \eqref{peaks2}, while the coefficient of variation obeys Eq. \eqref{peaks3} and tends to $\sqrt{2}$ as $M \to \infty$.
This means that the limiting form of the distribution of a single trajectory periodogram for such a choice of $f$
should be the $\chi^2$-distribution.
 We show below that this is indeed the case.

 For  $M$ divisible by $4$, 
 the kernel in \eqref{dis_phi} is  given by
\begin{align}
\label{1_05}
&\prod_{j=1}^{M}\cos\left(a_{j}x+b_{j}y\right) =  \cos^{\frac{M}{2}}(x)   =  \nonumber\\
& = \frac{1}{2^{\frac{M}{2}}} \sum_{m =-\frac{M}{4}}^{\frac{M}{4}} \binom{\frac{M}{2}}{\frac{M}{4} + m} \, \cos\left(2 m x\right) \,,
\end{align}
where $\binom{a}{b}$ is the binomial coefficient.
Substituting Eq. \eqref{1_05} into Eq.  \eqref{dis_phi}
 and integrating term by term, we find that for $f \Delta = \pi$ the moment-generating function 
 is given by the following sum of exponential functions:
\begin{align}
\label{1_07}
\Phi_{\lambda}\left(R_M\left(\frac{\pi}{\Delta}\right)\right) &= \frac{1}{2^{\frac{M}{2}}} \sum_{m=-\frac{M}{4}}^{\frac{M}{4}} \binom{\frac{M}{2}}{\frac{M}{4} + m} \, \nonumber\\
& \times \exp\left(-\frac{8D\Delta^{2} m^2}{M} \lambda \right) \, ,
\end{align}
and hence, the distribution of a single-trajectory periodogram is given explicitly as a sum of delta-functions 
\begin{align}
\label{1_08}
&P\left(R_{M}\left(\frac{\pi}{\Delta}\right)=R\right) = \frac{1}{2^{\frac{M}{2}}} \Bigg[ \binom{\frac{M}{2}}{\frac{M}{4}} \delta\left(R\right) + \nonumber\\
& + 2\sum_{m=1}^{\frac{M}{4}}\binom{\frac{M}{2}}{\frac{M}{4} + m} \delta\left(R-\frac{8D\Delta^{2}}{M} m^{2}\right) \Bigg]  \, .
\end{align}
Note that again the delta-peak at $R = 0$ is a consequence of the specific choice of $M$. For $M$ not divisible by $4$ such a peak is absent, precisely as it happens for a standard random walk, for which the parity of the number $M$ of steps matters.

In the limit $M \to \infty$ it is legitimate to replace the summation operation by an integration and use
\begin{align}
\label{1_09}
2^{-\frac{M}{2}} \binom{\frac{M}{2}}{m} \rightarrow \frac{2}{\sqrt{\pi M}} \textrm{e}^{-\frac{(m - M/4)^{2}}{M/4}} \, ,
\end{align}
to get from expression \eqref{1_07} the following limiting expression for the moment-generating function
\begin{align}
\label{1_10}
\Phi_{\lambda}\left(R_M\left(\frac{\pi}{\Delta}\right)\right) &\sim \int_{-M/4}^{M/4}\textrm{d}m \, \frac{2}{\sqrt{\pi M}} \textrm{e}^{-\left( 1 + 2D\Delta^{2}\lambda \right) \frac{4 m^{2}}{M}} \nonumber\\
&= \frac{\textrm{erf}\left( \sqrt{M(1+2D\Delta^{2}\lambda)}/2\right)}{\sqrt{1+2D\Delta^{2}\lambda}} \, . 
\end{align}
Recalling that $\mu_{R}(\pi/\Delta,T)=D\Delta^{2}$ we see then that for 
$M \to \infty$ the right-hand-side of the latter equation converges to
\begin{align}
\label{1_11}
\Phi_{\lambda}\left(R_M\left(\frac{\pi}{\Delta}\right)\right) \sim \left( 1+2 \mu_{R}\left(\frac{\pi}{\Delta},T\right) \, \lambda \right)^{-1/2} \, ,
\end{align}
which is precisely our conjectured expression \eqref{period} corresponding to $\gamma_R = \sqrt{2}$. The
distribution is then obtained upon inverting the Laplace transform to get
\begin{align}
\label{1_12}
P\left(R_M\left(\frac{\pi}{\Delta}\right)=R\right) &= \frac{1}{\sqrt{2 \pi \mu_{R}\left(\frac{\pi}{\Delta},T\right) \, R}} \nonumber\\
& \times \exp\left(-\frac{R}{2\mu_{R}\left(\frac{\pi}{\Delta},T\right)}\right)\, ,
\end{align}
which is the $\chi^2$-distribution conjectured in Eq. \eqref{dist10}. 
The convergence of Eq. \eqref{1_07} to \eqref{1_11}, and also of \eqref{1_08} to \eqref{1_12} for different values of $M$ is 
demonstrated in Fig. \ref{fig_plot1}.

\subsection{The case $f \Delta =\pi/2$}

Consider the case when $f T$ is at one-quarter of the prime period. In this case  
the coefficients $a_{j}$ and $b_{j}$ are periodic functions of $j$ with period $4$, i.e., $a_j = a_{j+4}$ and $b_j = b_{j+4}$. Within the period, their values are given by
\begin{equation}
\label{2_01}
\begin{tabular}{|c|c|c|c|c|}\hline
$j$ 		& 1 & 2 & 3 & 4 \\\hline
$a_{j}$ 	& 0 & 0 & 1 & 1 \\\hline
$b_{j}$ 	& 0 & -1 & -1 & 0 \\\hline
\end{tabular}
\end{equation}
such that random walk $W_M$ \eqref{WM} pauses at first two "time" moments
and steps on two consecutive ones. The random walk $V_M$ \eqref{VM} pauses at the first time moment, then makes two steps and pauses again. 

\begin{figure}
\includegraphics[width=8.5cm]{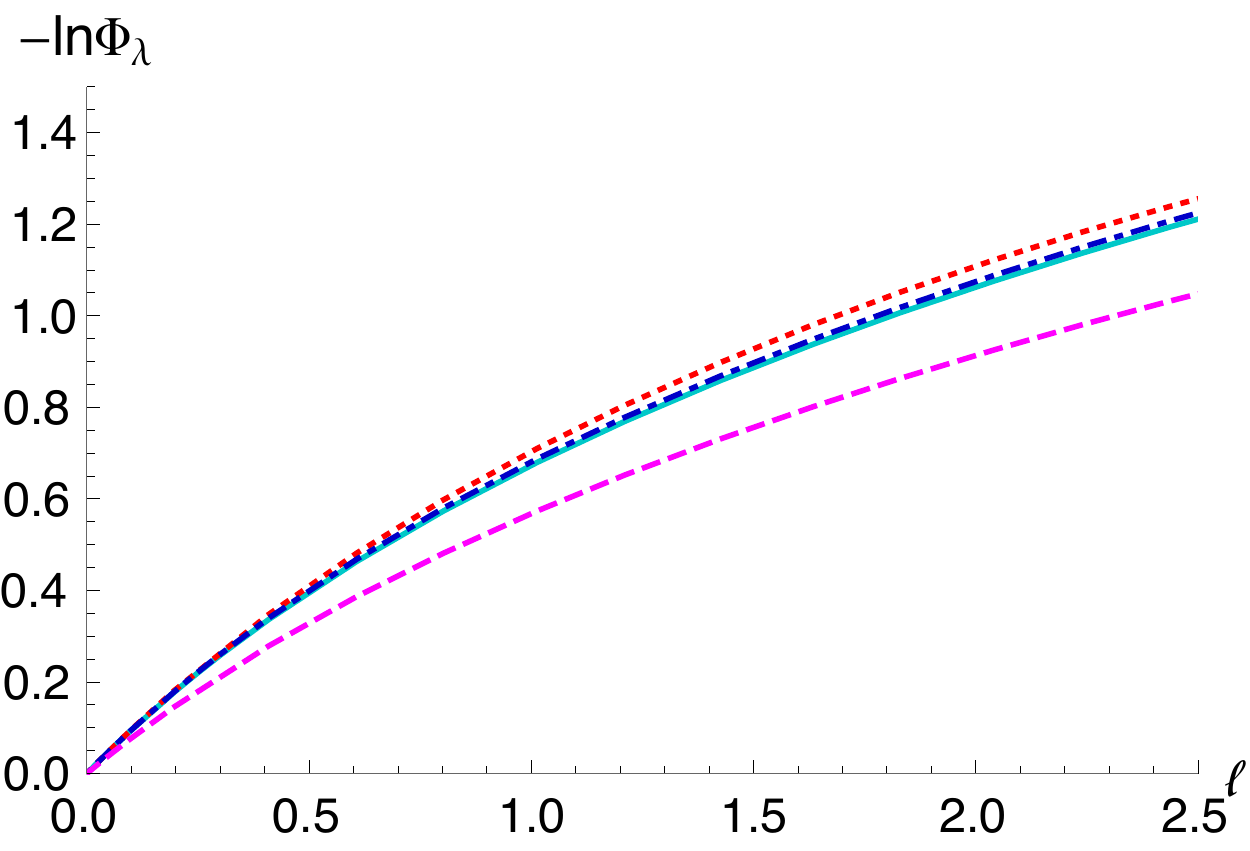}
\includegraphics[width=8.5cm]{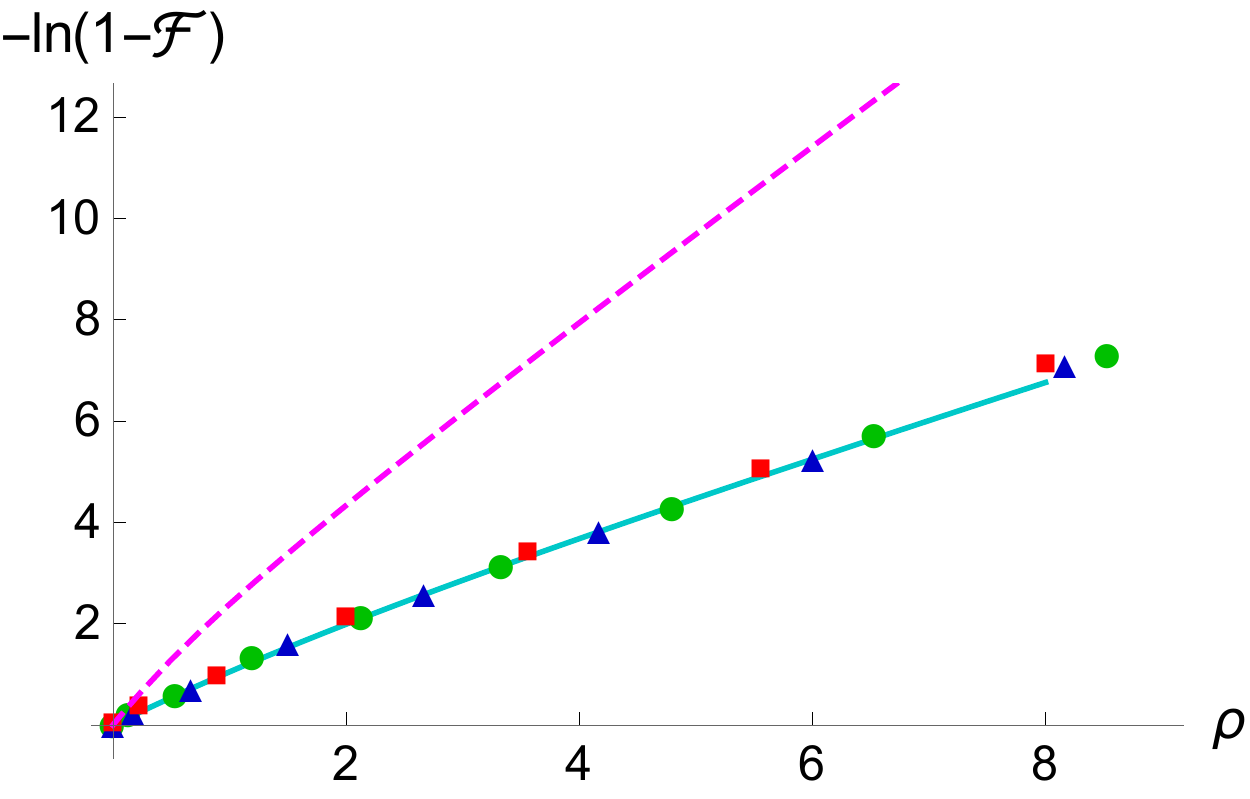}
\caption{The case $f \Delta = \pi/2$ (one-quarter of the prime period).
Upper panel: The solid curve is the conjectured moment-generating function in Eq. \eqref{period}. The dotted 
and the dotdashed curves give 
the exact moment-generating functions in the discrete-time case with  $M = 8$ and $M=16$, respectively.
The dashed line  depicts our continuous-time prediction in Eq. \eqref{MG} with $f = \pi/(2 \Delta)$ and $T = 1$.
 Lower panel: Solid curve depicts the cumulative distribution function of the conjectured 
 distribution in Eq. \eqref{dist10} with $\mu_R = 2 D \Delta^2$, $\gamma_R = \sqrt{5} \sqrt{1 - 12/(5 M)}$, 
 $D = 1/2$, $M = 12$ and $T = 1$. Symbols represent the exact cumulative function  of the distribution in the discrete-time case: squares correspond to $M= 36$, triangles to $M = 48$ and circles to $M = 60$.
\label{fig_plot2}
}
\end{figure} 

Now, we again assume that $M$ is divisible by $4$. Then, the kernel in \eqref{dis_phi} reduces to
\begin{align}
\label{2_03}
& \prod_{j=1}^{M}\cos\left(a_{j}x+b_{j}y\right) = \Bigg[ \cos x \cos y \cos (x-y) \Bigg]^{\frac{M}{4}} \nonumber\\
& = \sum_{m=0}^{\frac{M}{4}}\binom{\frac{M}{4}}{\frac{M}{4}-m} \left( \cos x \cos y \right)^{\frac{M}{2}-m} \left( \sin x \sin y \right)^{m} \, ,
\end{align}
such that the contributions dependent on $x$ and on $y$ appear in the factorized form.
Inserting the latter expansion into Eq.  \eqref{dis_phi}
 and integrating term by term, we get
\begin{align}
\label{2_04}
\Phi_{\lambda} = \sum_{m=0}^{\frac{M}{4}}\binom{\frac{M}{4}}{m} \, \mathcal{U}^{2}\left(\lambda;\frac{M}{4}+m,\frac{M}{4}-m\right)\, ,
\end{align}
where 
$\mathcal{U}$ is  defined by
\begin{align} 
\label{2_05}
\mathcal{U}\left(\lambda;m,n\right)  &=  \sqrt{\frac{M}{8 \pi D \Delta^2 \lambda}} \int_{-\infty}^{+\infty}\textrm{d} x \, \cos^m x  \sin^n x  \nonumber\\
& \times  \exp\left(-\frac{M}{8 D \Delta^2 \lambda} x^2\right) \, .
\end{align}
The function $\mathcal{U}\left(\lambda;m,n\right)$ is identically equal to zero for odd $n$. For even $n$, $n = 2 k$, we write
\begin{align}
& \cos^m x  \sin^{2 k} x =  \cos^m x \left(1 - \cos^2 x\right)^k = \nonumber\\
& = \sum_{p=0}^k (-1)^p \binom{k}{p} \cos^{2 p + m} x \,.
\end{align}
Further on, for even $m$, $m = 2 q$, the latter sum can be written down  
as a finite series of cosines
\begin{align}
& \cos^{2 q} x  \sin^{2 k} x =  \cos^{2 q} x \left(1 - \cos^2 x\right)^k = \nonumber\\
& = \sum_{p=0}^k \frac{(-1)^p}{4^{p+q}} \binom{k}{p} \sum_{l=-p-q}^{p+q} \binom{2 p + 2 q}{p+q+l} \cos\left(2 l x\right) \,,
\end{align}
such that $\mathcal{U}\left(\lambda;2 q, 2 k\right) $ is given explicitly by
\begin{align} 
\label{2_050}
\mathcal{U}\left(\lambda;2 q, 2 k\right)  &=  \sum_{p=0}^k \frac{(-1)^p}{4^{p+q}} \binom{k}{p} \sum_{l=-p-q}^{p+q} \binom{2 p + 2 q}{p+q+l} \nonumber\\
& \times \exp\left(- \frac{8 D \Delta^2 l^2}{M} \lambda\right)
\end{align}
For odd $m$, $m=2 q + 1$, we have
\begin{align}
 \cos^{2 q + 1} x  \sin^{2 k} x &=  \frac{1}{2} \sum_{p=0}^k \frac{(-1)^p}{4^{p+q}} \binom{k}{p} 
 \sum_{l=-p-q}^{p+q} \binom{2 p + 2 q}{p+q+l} \nonumber\\
 & \times \Big(\cos\left((2 l + 1) x\right) +  \cos\left((2 l - 1) x\right)\Big) \,,
\end{align}
which gives
\begin{align} 
\label{2_051}
\mathcal{U}\left(\lambda;2 q+1, 2 k\right)  &= \frac{1}{2} \sum_{p=0}^k \frac{(-1)^p}{4^{p+q}} \binom{k}{p} 
 \sum_{l=-p-q}^{p+q} \binom{2 p + 2 q}{p+q+l} \nonumber\\
 & \times \Bigg(\exp\left(- \frac{8 D \Delta^2 (2 l + 1)^2}{M} \lambda\right) + \nonumber\\
 &+ \exp\left(- \frac{8 D \Delta^2 (2 l - 1)^2}{M} \lambda\right) \Bigg) \,.
 \end{align}
Equations \eqref{2_04}, \eqref{2_050} and \eqref{2_051} define completely the moment-generating function 
of a single-trajectory periodogram in the case $f = \pi/(2 \Delta)$. The distribution function follows straightforwardly by 
expanding $\mathcal{U}^2$ in a series of exponentials and taking the inverse Laplace transform.
 
In Fig.\ref{fig_plot2} we plot our exact result for the moment-generating function in Eq. \eqref{2_04} and the corresponding result for the distribution function of a single-trajectory periodogram, together with the conjectured expressions \eqref{period} and \eqref{dist10} with $\mu_R(f,T)$ and $\gamma_R$ defined by Eqs. \eqref{14} and  \eqref{zu}. We observe a very convincing convergence of the exact results, upon an increase of $M$, to the conjectured forms.

\subsection{The case $f \Delta= 2 \pi/3$}

In this case the coefficients $a_j$ \eqref{aj} and $b_j$ \eqref{bj}
are periodic functions of $j$ with period $3$, and are given by
\begin{equation}
\label{4_01}
\begin{tabular}{|c|c|c|c|}\hline
$j$ 		& 1 & 2 & 3 \\\hline
$a_{j}$ 	& 0 & $1/2$ & 1 \\\hline
$b_{j}$ 	& 0 & $-\sqrt{3}/2$ & 0 \\\hline
\end{tabular}
\end{equation}
within their period. 

\begin{figure}
\includegraphics[width=8.5cm]{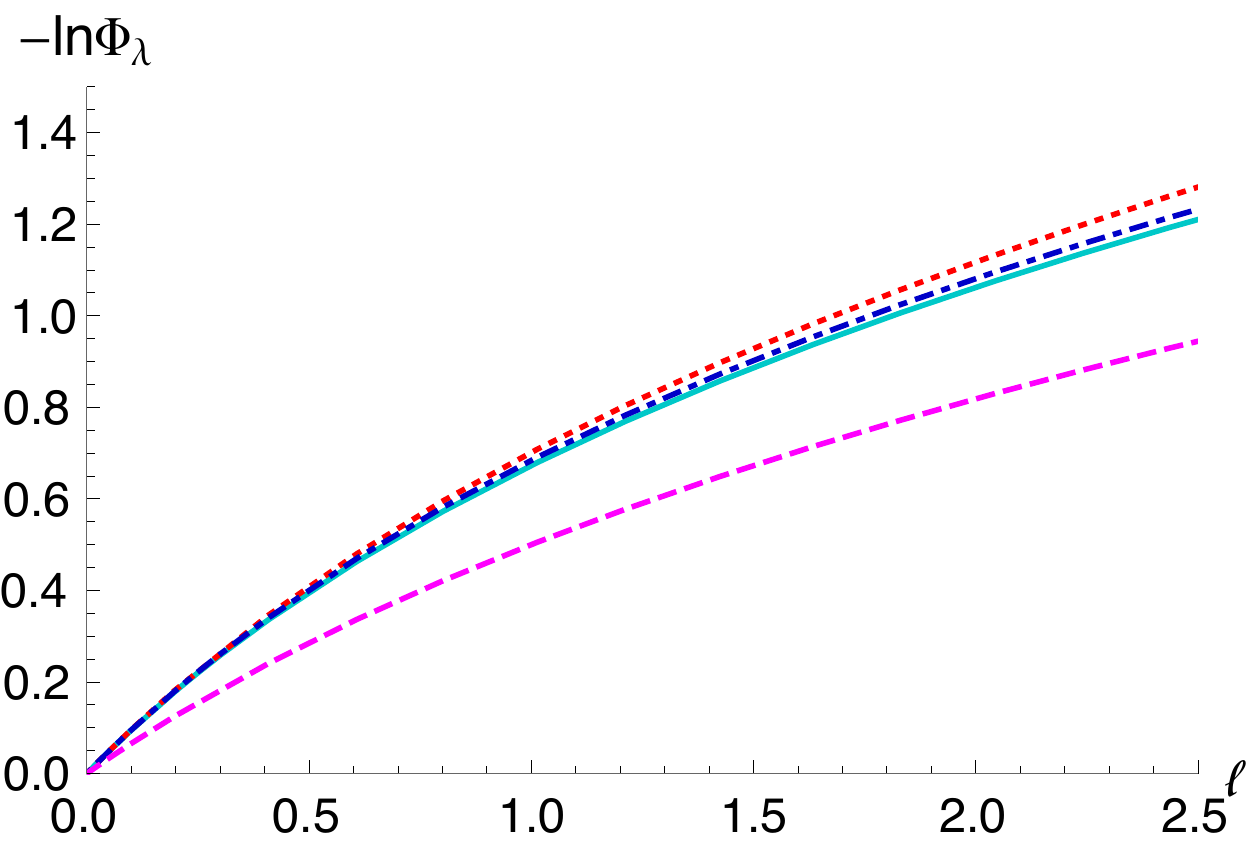}
\includegraphics[width=8.5cm]{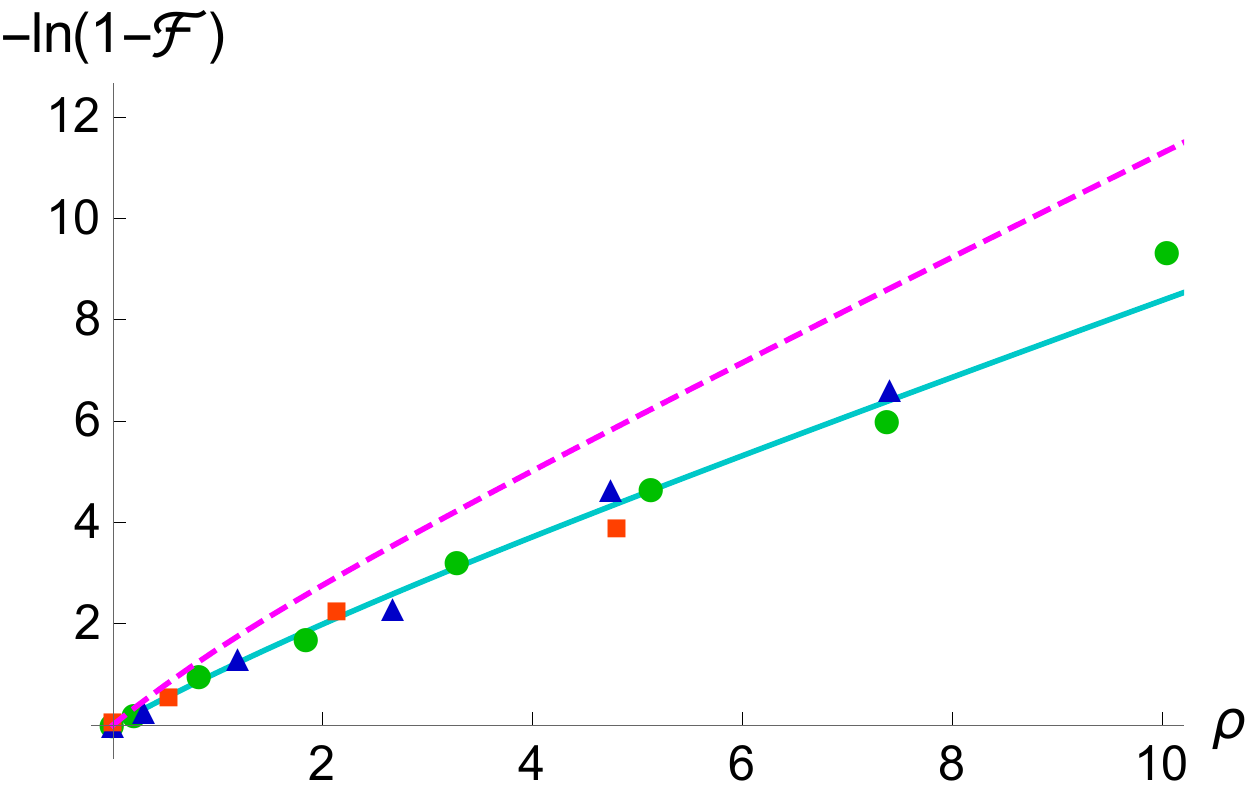}
\caption{The case $f \Delta = 2 \pi/3$ (one-third of the prime period).
Upper panel: The exact discrete-time 
moment-generating function in Eq. \eqref{v} versus ${\it l} = \lambda/\mu_R$ for $M = 9$ (dotted curve) and $M = 15$ (dotdashed curve).
The solid curve is the conjectured form in Eq. \eqref{period}. The dashed curve is our continuous-time prediction in \eqref{MG}. 
Lower panel: the cumulative distribution function ${\cal F}$ versus $\rho = R/\mu_R$. The solid curve is ${\cal F}$ associated with the conjectured form in Eq. \eqref{dist10}. Symbols depict the cumulative distribution function of the exact discrete-time distribution generated by Eq. \eqref{v}: squares correspond to $M= 15$, triangles to $M = 27$ and circles to $M = 39$. The dashed line is ${\cal F}$ for the continuous-time result in Eq.\eqref{dist}.
\label{fig_plot3}
}
\end{figure} 

We assume now that $M$ is divisible by $3$. In this case the kernel
\begin{align}
\label{4_02}
&\prod_{j=1}^{M}\cos\left(a_{j}x+b_{j}y\right) = \left( \cos x \cos\left(\frac{x-\sqrt{3}y}{2}\right) \right)^{\frac{M}{3}}  =  \cos^{\frac{M}{3}} x \nonumber\\
& \times  \sum_{p=0}^{\frac{M}{3}} \binom{\frac{M}{3}}{p} \left(\cos \frac{x}{2} \cos \frac{\sqrt{3} y}{2}\right)^p \left(\sin \frac{x}{2} \sin \frac{\sqrt{3} y}{2}\right)^{\frac{M}{3}-p}
\end{align}
Next, we use
\begin{align}
&\cos^{\frac{M}{3}} x = \left(2 \cos^2 \frac{x}{2} - 1\right)^{\frac{M}{3}} = \nonumber\\
&= \sum_{k=0}^{\frac{M}{3}} (-1)^k \binom{\frac{M}{3}}{k} \left(2 \cos^2 \frac{x}{2}\right)^{\frac{M}{3}-k}
\end{align}
which gives, in combination with \eqref{4_02}, the following representation of the kernel
\begin{align}
&\prod_{j=1}^{M}\cos\left(a_{j}x+b_{j}y\right) = 2^{\frac{M}{3}}\sum_{k=0}^{\frac{M}{3}} \frac{(-1)^k}{2^k} \binom{\frac{M}{3}}{k} \sum_{p=0}^{\frac{M}{3}} \binom{\frac{M}{3}}{p} \nonumber\\
& \times \cos^{p+ \frac{2 M}{3} - 2 k}\frac{x}{2} \sin^{\frac{M}{3}-p}\frac{x}{2} \cos^{p} \frac{\sqrt{3} y}{2} \sin^{\frac{M}{3} - p}  \frac{\sqrt{3} y}{2} \,.
\end{align}
Inserting the latter expansion into Eq.  \eqref{dis_phi}
 and integrating term by term, we obtain the following exact result for the moment-generating function of a single-trajectory periodogram 
 \begin{align}
 \label{v}
& \Phi_{\lambda} = 2^{\frac{M}{3}}\sum_{k=0}^{\frac{M}{3}} \frac{(-1)^k}{2^k} \binom{\frac{M}{3}}{k} \sum_{p=0}^{\frac{M}{3}} \binom{\frac{M}{3}}{p} \nonumber\\
&\times \mathcal{U}\left(\frac{\lambda}{4}; p + \frac{2 M}{3} - 2k, \frac{M}{3} - p\right) \,  \mathcal{U}\left(\frac{3 \lambda}{4}; p, \frac{M}{3} - p\right)  \,,
 \end{align}
 where $\mathcal{U}$ is defined in Eqs. \eqref{2_050} and \eqref{2_051}. The moment-generating function is a finite sum of exponentials and the corresponding distribution is a sum of delta-functions with coefficients which can be readily deduced from Eq. \eqref{v}. In Fig. \ref{fig_plot3} we demonstrate that $\Phi_{\lambda}$ in Eq. \eqref{v}, and also the cumulative distribution function of the distribution deduced from Eq. \eqref{v}, do indeed converge to the conjectured forms \eqref{period} and \eqref{dist10} with $\mu_R(2 \pi/(3 \Delta),T)$ and $\sigma^2_R( 2 \pi/(3 \Delta),T)$, Eqs. \eqref{onethird1} and \eqref{onethird2}.

\subsection{The case $f \Delta =  2 \pi$}

We finally consider the special case when $f T$ is at the end-point of the
prime period. In this case, $a_j = M + 1 - j$ and $b_j \equiv 0$, such that we are led to the special case ($f =0$) 
studied already in Appendix \ref{A}. Convergence of the discrete-time moment-generating function in Eq. \eqref{MGdisf0}
and the cumulative distribution function of the distribution in Eq. \eqref{DISTdiscf0} to the conjectured forms in Eqs. \eqref{period} and \eqref{dist10} is evident.

\end{document}